\documentstyle[12pt,aaspp4,epsf]{article}

\def\bs{\hbox{\boldmath$\sigma$}}

\begin{document}
\title{Supernova Neutrino Opacity from Nucleon-Nucleon
Bremsstrahlung and Related Processes}

\author{Steen Hannestad}
\affil{Max-Planck-Institut f\"ur Physik (Werner-Heisenberg-Institut)\\
F\"ohringer Ring 6, 80805 M\"unchen, Germany\\
and Theoretical Astrophysics Center, University of Aarhus,
DK-8000 Aarhus C, Denmark}

\author{Georg Raffelt}
\affil{Max-Planck-Institut f\"ur Physik (Werner-Heisenberg-Institut)\\
F\"ohringer Ring 6, 80805 M\"unchen, Germany}

\date{\today}

\begin{abstract}
  Elastic scattering on nucleons, $\nu N\to N\nu$, is the dominant
  supernova (SN) opacity source for $\mu$ and $\tau$ neutrinos. The
  dominant energy- and number-changing processes were thought to be
  $\nu e^-\to e^-\nu$ and $\nu\bar\nu\leftrightarrow e^+e^-$ until
  Suzuki (1993) showed that the bremsstrahlung process $\nu\bar\nu
  NN\leftrightarrow NN$ was actually more important.  We find that for
  energy exchange, the related ``inelastic scattering process'' $\nu
  NN\leftrightarrow NN \nu$ is even more effective by about a factor
  of 10. A simple estimate implies that the $\nu_\mu$ and $\nu_\tau$
  spectra emitted during the Kelvin-Helmholtz cooling phase are much
  closer to that of $\bar\nu_e$ than had been thought previously.  To
  facilitate a numerical study of the spectra formation we derive a
  scattering kernel which governs both bremsstrahlung and inelastic
  scattering and give an analytic approximation formula.  We consider
  only neutron-neutron interactions, we use a one-pion exchange
  potential in Born approximation, nonrelativistic neutrons, and the
  long-wavelength limit, simplifications which appear justified for
  the surface layers of a SN core.  We include the pion mass in the
  potential and we allow for an arbitrary degree of neutron
  degeneracy.  Our treatment does not include the neutron-proton
  process and does not include nucleon-nucleon correlations. Our
  perturbative approach applies only to the SN surface layers, i.e.\
  to densities below about $10^{14}\,\rm g\,cm^{-3}$.
\end{abstract}

\newpage


\section{Introduction}

A quantitatively accurate prediction of the fluxes and spectra of
supernova (SN) neutrinos is required for a theoretical understanding
of an array of fascinating phenomena, including the explosion
mechanism itself, the cause of neutron star natal kicks, various
aspects of SN nucleosynthesis, and the interpretation of the neutrino
signal from SN~1987A and future galalactic SNe, with or without the
assumption of neutrino masses and mixings. However, at the present
time one is far away from this goal because of numerical limitations
and because of significant shortcomings in the calculation of the
relevant microphysics, notably the equation of state and the neutrino
opacities. 

We presently study the opacity contribution of nucleon bremsstrahlung
$NN\to NN\nu\bar\nu$ and its inverse $\nu\bar\nu NN\to NN$, processes
which have been ignored in all numerical SN studies except for the
proto-neutron star cooling calculations of Suzuki (1991, 1993) who
found a significant modification of the neutrino fluxes and
spectra. We also study the inelastic scattering process $\nu N N\to
NN\nu$ which is obtained when crossing a final-state bremsstrahlung
neutrino into the initial state. Janka et~al.\ (1996) had stressed its
apparent importance for the neutrino spectra formation, but without
making a connection with Suzuki's work.  We will again motivate the
importance of these processes and provide a scattering kernel which
allows for a practical implementation in a numerical SN code.  While
this is a modest aspiration relative to the large number of open
questions regarding SN neutrino opacities, we hope that our work may
nevertheless prove useful for future numerical studies.

In order to appreciate the importance of nucleon bremsstrahlung and
related reactions recall that for $\nu_e$ and $\bar\nu_e$ the dominant
opacity contribution derives from charged-current processes ($\beta$
processes) which involve electrons and positrons. In the diffusion
regime they keep $\nu_e$ and $\bar\nu_e$ essentially in local thermal
equilibrium. We are mostly concerned, however, with the transport of
$\nu_\mu$, $\nu_\tau$, and their antiparticles, to which we will
collectively refer as $\nu_\mu$. The dominant opacity contribution
is the neutral-current scattering on nucleons $\nu_\mu N\to N\nu_\mu$,
a process which does not equilibrate the neutrino number density and
which is ineffective at modifying the spectrum because the neutrino
energies are low relative to the mass of the nonrelativistic nucleons,
especially in the outer layers of a SN core. Therefore, processes such
as $\nu_\mu e^-\to e^-\nu_\mu$ and $\nu_\mu\bar\nu_\mu\leftrightarrow
e^+e^-$, which are subdominant with regard to the total opacity, are
nevertheless important for the equilibration of the neutrino number
density and spectra. It turns out that nucleon bremsstrahlung is far
more effective than pair annihilation at equilibrating the neutrino
number density, and it is of comparable importance to $\nu_\mu e^-\to
e^-\nu_\mu$ at equilibrating the spectra. Moreover, the inelastic
scattering process $\nu NN\to NN\nu$ is in turn far more effective at
exchanging energy than bremsstrahlung and is thus the dominant process
for the neutrino spectra formation.  Including these processes in a
proto-neutron star cooling calculation would make the $\nu_\mu$
spectrum far more similar to that of $\bar\nu_e$ than had been
thought.

While nucleon bremsstrahlung and related processes are conceptually
simple, a reliable calculation is nevertheless nontrivial.
Difficulties include the nuclear matrix element, i.e.\ the appropriate
nucleon-nucleon interaction potential, the intermediate degree of
nucleon degeneracy, $NN$ correlations, and the role of
multiple-scattering effects.  We will not be able to resolve all of
these issues. For example, we will completely ignore $NN$ correlations
which may be quite important in some regions of the SN core where
nuclei may not even be completely dissociated. In regions where nuclei
exist, nuclear bound-bound or bound-free transitions involving
neutral-current neutrino reactions may be important. The main advance
of our calculation is the treatment of intermediate degrees of nucleon
degeneracy in free-free transitions, the inclusion of the pion mass in
the nucleon interaction potential, and the inclusion of
multiple-scattering effects. Our scattering kernel is then
self-consistent in the sense that it allows for the simultaneous
treatment of bremsstrahlung and inelastic scattering while producing
the correct total $\nu N$ scattering cross section.

We focus on the neutrino spectra formation which takes place in the
surface layers of a proto neutron star where number-equilibrating,
energy-equilibrating, and finally scattering processes freeze out at
different radii so that the different neutrino species are emitted
with different fluxes and different spectral temperatures, but
approximately equal luminosities.  We expect that bremsstrahlung and
inelastic scattering make the spectra and fluxes more similar than had
been thought previously.

Nucleon-nucleon interactions not only lead to bremsstrahlung and the
inelasticity of neutrino-nucleon scattering, but also to a reduction
of the overall $\nu N$ scattering cross section (Raffelt \& Seckel
1995; Sawyer 1995; Raffelt, Seckel \& Sigl 1996). This cross-section
reduction is an inevitable consequence of the fact that nucleon spins
interact so that one cannot consistently include bremsstrahlung and
the inelasticity of $\nu N$ scattering and yet ignore this
cross-section reduction.  Unfortunately, for densities above, say,
$10^{14}\,\rm g\,cm^{-3}$ a controlled calculation of the scattering
kernel and thus of the neutrino transport coefficients is currently
out of reach.  The only hint as to the possible magnitude of the
cross-section reduction in the inner parts of a SN core derives from
the signal duration of SN~1987A (Keil, Janka \& Raffelt 1995) and the
$f$-sum rule of the spin-density structure function (Sigl 1996) which
imply that a naive extrapolation of the perturbative spin-density
structure function into the high-density regime would significantly
overestimate the cross-section suppression.  Recent 2-dimensional
hydrodynamic SN cooling calculations indicate that actually
convection, not neutrino radiative transport, may be the dominant mode
of energy transfer from the inner proto-neutron star to the neutrino
sphere (Keil, Janka \& M\"uller 1996), rendering the high-density
opacities less critical for the overall neutrino luminosity. In any
case, the problem of the neutrino spectra formation in the surface
layers is rather disjoint from the problem of calculating the overall
neutrino luminosity.

We begin our discussion in Sec.~2 with a comparison of the different
neutrino opacity contributions and thus motivate the importance of
nucleon bremsstrahlung and related processes.  In Sec.~3 we formulate
the bremsstrahlung rate in terms of the dynamical spin-density
structure function (the scattering kernel) and thus reduce the problem
to the calculation of a dimensionless function which includes all of
the nuclear and many-body complications.  We also discuss the
underlying nuclear matrix element of the bremsstrahlung process.  In
Sec.~4 we study the nucleon phase space and provide the appropriate
analytic approximation formulae. In Sec.~5 we summarize and discuss
our results.


\section{Comparison of Opacity Sources}

\subsection{Mean Free Path}

In a SN core the main opacity source for $\mu$ and $\tau$ neutrinos is
neutral-current scattering on nucleons.  However, the large nucleon
mass relative to typical neutrino energies renders this process
ineffective at equilibrating the neutrino spectra and it certainly
cannot modify the neutrino number density.  We begin with a comparison
of the relative importance of those processes which are subdominant
with regard to the total opacity, yet allow for an effective
modification of the $\nu_\mu$ spectrum and density.  The simplest case
is $\nu_\mu\bar\nu_\mu\to\nu_e\bar\nu_e$. Using Maxwell-Boltzmann
distributions at temperature $T$ for all neutrino species, we find the
thermally averaged absorption rate
\begin{equation}
\Gamma_0\equiv\frac{4}{\pi^3} \, G_F^2 T^5,
\end{equation}
which sets a natural scale for all other processes.  We use natural
units with $\hbar=c=k_{\rm B}=1$.  The rate for
$\nu_\mu\nu_e\to\nu_\mu\nu_e$ together with that for
$\nu_\mu\bar\nu_e\to\nu_\mu\bar\nu_e$ is $4\Gamma_0$. For
$\nu_\mu\bar\nu_\mu\to e^+e^-$ it is
$\Gamma_0\,(C_{V,e}^{2}+C_{A,e}^{2})\eta_e^{4} e^{-\eta_e}/12$ while
it is $\Gamma_0\,(C_{V,e}^{2}+C_{A,e}^{2})3 \eta_e^2$ for $\nu_{\mu}
e^{\pm} \to \nu_{\mu} e^{\pm}$, where $\eta_e\gg1$ is the electron
degeneracy parameter.  The weak coupling constants are
$C_{A,e}=-\frac{1}{2}$ and $C_{V,e}=-\frac{1}{2}+2\sin^2\Theta_{\rm
W}$ for $\nu_{\mu}$ and $\nu_{\tau}$.  Neutrino coalescence
$\nu\bar{\nu}\to{\rm plasmon}$ is negligible.  Finally, there is
inverse bremsstrahlung on nucleons $\nu_{\mu}\bar{\nu}_{\mu}NN\to NN$;
the spectrally averaged absorption rate is given by Eq.~(\ref{simple})
in Sec.~3.4.

In order to compare these different processes for realistic conditions
we use the numerical SN model S2BH\_0 of Keil, Janka \& Raffelt (1995)
which represents a SN core $1\,\rm s$ after collapse; the radial
profiles of various physical parameters are shown in Fig.~1.  In the
outer parts of the star, which is where the neutrino spectra are
formed, the effective nucleon degeneracy parameter $\eta_*=p_{F}^2/2
m_N T$ varies between 2 and 4, $\eta_e$ between 1 and 8, and the range
of relevant temperatures is taken to be $T=5\hbox{--}10~{\rm MeV}$,
leading to
\begin{equation}\label{mfplist}
\Gamma=\Gamma_0\times\cases{
0.03\hbox{--}0.4&for $\nu_{\mu} \bar{\nu}_{\mu} \to e^+ e^-$,\cr
1&for $\nu_{\mu}\bar{\nu}_{\mu}\to\nu_{e} \bar{\nu}_{e}$,\cr 
4&for $\nu_{\mu}{\nu}_{e}\to\nu_{\mu}\nu_{e}$
plus $\nu_{\mu}\bar{\nu}_{e}\to\nu_{\mu}\bar{\nu}_{e}$,\cr
2\hbox{--}50&for $\nu_{\mu}e^{+}\to\nu_{\mu}e^{+}$ plus
$\nu_{\mu}e^{-}\to\nu_{\mu}e^{-}$,\cr
15\hbox{--}300&for $\nu_{\mu} \bar{\nu}_{\mu}nn \to nn$.}
\end{equation}
We conclude that for typical SN conditions bremsstrahlung is by far
the most important number-changing reaction. Even for energy exchange
it is more important than elastic scattering on electrons and
positrons.

\subsection{Energy Transfer}

The processes considered in the previous section are very effective at
transferring energy in the sense that in a given interaction the
neutrino essentially loses all memory of its previous energy.  This is
not the case for the ``inelastic scattering process'' $\nu n n\to nn
\nu$ which is the bremsstrahlung process with a final-state neutrino
crossed into the initial state. While we use the term ``inelastic
scattering process'' we stress that it is not logically distinct from
the ``elastic channel'' $\nu n\to n\nu$. Both are described by one and
the same scattering kernel which has a finite width as a function of
the energy transfer $\omega$ due to the nucleon-nucleon
interaction. Therefore, in each collision the final-state neutrino
energy is smeared out by a small amount which is given by the width of
the scattering kernel.

To quantify the efficiency of energy transfer of this process relative
to those of the previous section it is useful to consider a fluid of
neutrinos at a temperature $T_\nu$ which is slightly different than
the temperature $T$ of the neutron bath.  According to
Eqs.~(\ref{bremstransfer}) and (\ref{scatteringtransfer}) we find for
the rate of energy transfer between these fluids
\begin{equation}
\frac{\Delta Q}{\Delta T}=
\frac{3 C_{A}^2 G_F^2 n_B T^5}{\pi^3}
\,\frac{\Gamma_\sigma}{2\pi T}
\times\cases{1&for $nn\leftrightarrow nn\nu\bar\nu$,\cr
20&for $\nu nn\leftrightarrow nn\nu$,\cr}
\end{equation}
where the spin-fluctuation rate $\Gamma_\sigma$ is given in
Eq.~(\ref{eq:gamsig}). For the bremsstrahlung case we have included a
factor $1/2$ relative to Eq.~(\ref{bremstransfer}); we only count the
energy transferred to the neutrinos.  These results receive numerical
corrections of order unity for realistic conditions, reducing the
relative importance of scattering to perhaps a factor of~10.

For completeness we have calculated the same quantity for 
neutrino scattering on degenerate electrons, 
\begin{equation}
\frac{\Delta Q_{\nu e^-\leftrightarrow e^-\nu}}{\Delta T}=
\frac{24}{\pi^5} \, G_F^2 (C_{V,e}^2+C_{A,e}^2) \eta_e^2 T^8,
\end{equation}
where again $C_{V,e}\approx0$ and $C_{A,e}=-\frac{1}{2}$ for 
$\nu_\mu$. Relative to inelastic nucleon scattering this is
\begin{equation}
\frac{\Delta Q_{\nu_\mu e^-\leftrightarrow e^-\nu_\mu}}
{\Delta Q_{\nu_\mu nn\leftrightarrow nn\nu_\mu}}=
1.37 \times 10^{-3} \, \frac{\eta_e^2}{\eta_{\ast}^3}T_{10}^{1/2},
\end{equation}
where $T_{10}=T/10~{\rm MeV}$. In the outer part of our SN
model we have $\eta_e<8$, $\eta_*=2$--4, and $T_{10}=0.5$--1 so that
the ratio is always smaller than $0.01$.  Electrons are negligible for
the energy transfer, in complete keeping with the conclusions of Janka
et~al.~(1996).

In the calculation of bremsstrahlung and related processes we assume
nonrelativistic nucleons and use the long-wavelength approximation. As
a consequence neutrinos cannot transfer energy to the nucleons if the
$NN$ interaction is switched off. From Janka (1991) and Tubbs (1979)
and using a Maxwell-Boltzmann distribution for the neutrinos we find
\begin{equation}
\frac{\Delta Q_{\rm recoil}}{\Delta T}=
\frac{(C_{V}^2+5 C_{A}^2) G_F^2 n_B T^5}{\pi^3}
\,\frac{240\,T}{m_N}.
\end{equation}
Naturally, the vector-current interaction appears here in contrast to
bremsstrahlung-related effects. Ignoring $C_{V}^2$ we find
\begin{equation}
\frac{\Delta Q_{\nu nn\leftrightarrow nn\nu}}{\Delta Q_{\rm recoil}}=
\frac{m_N \Gamma_\sigma}{40\pi T^2}
=\alpha_\pi^2\,\frac{2\sqrt{2\pi}}{15\,\pi^3}\,\eta_*^{3/2}
=1.21\,\eta_*^{3/2},
\end{equation}
where we have used Eq.~(\ref{eq:gamsig}) for the spin-fluctuation
rate. The vector current contribution to recoils reduces this ratio a
bit, and pion-mass effects reduce the inelastic term by another small
amount so that it may be more realistic to take something like
$0.5\,\eta_*^{3/2}$. In the SN core model of Fig.~1 we always have
$\eta_*=2$--4 so that recoil effects are always of roughly equal
importance to inelastic scattering.  Therefore, the transfer of energy
is even more effective than what is accounted for by our scattering
kernel which ignores nucleon recoils.

\subsection{Energy Sphere}

We may next attempt to estimate the change in the spectral temperature
of the emitted neutrino fluxes which is brought about by including
bremsstrahlung and related processes. Neutrinos stream off freely from
the ``transport sphere'' which is the radius where scattering on
nucleons is no longer effective at trapping them. Deeper inside is the
``energy sphere'' which is the radius where energy-exchanging
processes such as scattering on electrons become ineffective (Burrows
\& Mazurek 1982, 1983).  We presently estimate the temperature of the
energy sphere for the SN model of Fig.~1 with or without the inclusion
of bremsstrahlung and inelastic scattering.  We take the change in
this temperature as an indication of the changed spectral temperature
of the emitted neutrinos. We stress, of course, that the temperature
at the energy sphere is not identical with the spectral temperature of
the emitted neutrino flux which is modified by the fact that lower
energy neutrinos diffuse more effectively through the layer between
the energy and the transport spheres.  Still, we think that the
following procedure gives one a sense of the quantitative importance
of bremsstrahlung and its sister reactions.

In order to calculate the location of the energy sphere we introduce
the effective mean free path for energy exchange
(Shapiro \& Teukolsky 1983)
\begin{equation}
\lambda_{\rm eff}^{-1}=
\sqrt{\lambda_{\rm tot}^{-1}\lambda_{\rm e}^{-1}}.
\end{equation}
Here, $\lambda_{\rm tot}$ is the mean free path for a neutrino of a
given energy while $\lambda_{\rm e}$ is the mean free path against
energy-changing reactions. If energy-changing reactions dominate the
total opacity we have $\lambda_{\rm tot}\simeq\lambda_{\rm e}$ and
thus $\lambda_{\rm eff}\simeq\lambda_{\rm e}$. However, in a
``scattering atmosphere'' as in a SN core where $\lambda_{\rm
tot}\simeq\lambda_{\rm scatter}$ the trapping of neutrinos by
collisions gives them a larger chance to lose energy by some other
process, leading to an increased value of $\lambda_{\rm eff}^{-1}$
relative to $\lambda_{\rm e}^{-1}$. We define the radius $R_{\rm
e}$ of the energy sphere by
\begin{equation}\label{energysphere}
\int_{R_{\rm e}}^\infty dr\,\lambda_{\rm eff}^{-1}=
{\textstyle \frac{2}{3}},
\end{equation}
i.e.\ by the radius where the effective optical depth is
$\frac{2}{3}$.

We have performed this calculation for the SN model of Fig.~1 first
for muon neutrinos by including the neutral-current scattering on
protons and neutrons as the dominant scattering process. Nucleon
degeneracy effects were taken into account. As an energy-exchange
process we first use electron scattering, again including degeneracy
effects. We then find that the temperature of the energy sphere varies
with neutrino energy as shown by the solid line in Fig.~2. Next we use
inverse bremsstrahlung as the only energy-changing reaction according
to the numerical prescription developed in Sec.~4, leading to the
short-dashed line. We see that the two processes are of roughly equal
importance, with electron scattering dominating for relatively large
neutrino energies.  We have repeated the same exercise for
$\bar\nu_e$, except that we include the charged-current reaction
$\bar\nu_e p\to n e^+$ as an energy-changing reaction, leading to the
long-dashed curve. Evidently this is the most important reaction for
$\bar\nu_e$ and thus dominates the spectrum formation.

To extract one fixed energy-sphere radius we next take neutrinos to be
in thermal equilibrium up to their energy sphere so that their
distribution is characterized by the local medium temperature, and
that further out they are characterized by their energy-sphere
temperature. All reaction rates are averaged with this spectral
distribution so that
\begin{equation}
\langle \lambda_{\rm eff}^{-1}\rangle
=\sqrt{\langle \lambda_{\rm tot}^{-1}\rangle
\langle \lambda_{\rm e}^{-1}\rangle}.
\end{equation}
The energy sphere is defined as in Eq.~(\ref{energysphere}) except
using $\langle \lambda_{\rm eff}^{-1}\rangle$ instead of $\lambda_{\rm
eff}^{-1}$. In order to locate the neutrino sphere we must now perform
an iteration, assuming first some estimate for the energy-sphere
temperature to determine the location where the effective optical
depth is $\frac{2}{3}$, then use the temperature there as the next
approximation, and so forth until convergence (Keil \& Janka 1995).
We stress that our benchmark model was calculated with equilibrium
neutrino transport everywhere so that it becomes actually more
self-consistent, not less so, when the neutrinos are kept in
equilibrium out to larger radii.

For the SN core model of Fig.~1 we find the energy-sphere temperatures
and densities shown in Table~1. We have included the bremsstrahlung
process according to the prescription developed in Secs.~3 and 4, and
then multiplied its rate by a factor $f_{\rm brems}$. For $f_{\rm
brems}=0$ we find the energy sphere without bremsstrahlung while
$f_{\rm brems}=1$ includes the full effect.  We also show the
redshifted energy-sphere temperatures for an observer at infinity and
the total optical depth at the location of the energy sphere, again
for a spectrally averaged neutrino mean free path.  We are thus led to
conclude that bremsstrahlung alone reduces the $\nu_\mu$ spectral
temperature by about $1\,\rm MeV$, while that of $\bar\nu_e$ remains
largely unchanged. The change of the $\nu_\mu$ spectrum agrees
with Suzuki's (1991, 1993) results.

In order to estimate the impact of inelastic scattering we recall that
it is by a factor of 10 more important than bremsstrahlung, give or
take a factor of 2. Therefore, we have also calculated the energy
sphere for $f_{\rm brems}=5$, 10, and 20, which probably mimics the
effect of inelastic scattering and its uncertainty. The $\bar\nu_e$
temperature is still only mildly affected, suggesting that $\beta$
processes are still very important for this species.  The $\nu_\mu$
temperature is significantly lowered; it is very close to that of
$\bar\nu_e$.

Of course, including $NN$ interactions consistently will probably
accelerate the cooling process due to the reduced overall neutrino
scattering cross section in the deep interior of the SN core.  This
will likely heat the neutrino sphere so that the spectral temperatures
may actually {\it increase}. A similar effect will be caused by
convection if it is a generic phenomenon for the proto-neutron star
evolution.  We believe, however, that our schematic treatment gives us
a reasonable estimate of the {\it differential\/} temperature change,
i.e.\ the $\nu_\mu$ and $\bar\nu_e$ temperatures will be much closer
than is often assumed.

A quantitative assessment of the spectral modifications requires
detailed numerical simulations which we are in no position to
perform. In the following, however, we provide a scattering kernel
which is needed for such an investigation.


\section{Bremsstrahlung and Related Processes}

\subsection{Scattering Kernel}

For a numerical study the bremsstrahlung process, its inverse, and
inelastic scattering are best formulated in terms of a ``scattering
kernel'' which contains all of the properties of the nuclear medium,
but which excludes the neutrino phase-space integration. We begin with
the usual weak interaction Hamiltonian density\footnote{In previous
papers (e.g.\ Janka et al.\ 1996) it had been written
$(G_F/2\sqrt2)\ldots$ so that, for example, $C_A$ was $\pm 1.26$
rather than $\pm1.26/2$ which we use here.}
\begin{equation}
{\cal H}_{\rm int}=\frac{G_F}{\sqrt2}\,
\bar\psi_N\gamma_\mu(C_V-C_A\gamma_5)\psi_N
\bar\psi_\nu\gamma_\mu(1-\gamma_5)\psi_\nu 
\end{equation}
where $G_F$ is Fermi's constant, $\psi_N$ the nucleon Dirac field for
either protons or neutrons, and $\psi_\nu$ the neutrino field. The
neutral-current vector coupling constant is
$\frac{1}{2}-2\sin^2\Theta_{\rm W}\approx 0$ for protons and
$-\frac{1}{2}$ for neutrons. The axial-current coupling is often taken
to be $\pm 1.26/2$ for protons and neutrons, respectively, but the
strange-quark contribution to the nucleon spin causes certain
deviations from this simple picture (Raffelt \& Seckel 1995).

We are concerned with conditions where the temperature is below
$10\,\rm MeV$ so that a typical nucleon velocity is around
$0.2\,c$. Therefore, it is reasonable to use the nonrelativistic limit
which implies that only the axial vector current contributes to
bremsstrahlung (Friman \& Maxwell 1979; Raffelt \& Seckel 1995).  The
squared matrix element for the emission of a neutrino pair,
$N_1+N_2\to N_3+N_4+\nu+\bar\nu$, may be written as
\begin{equation}\label{eq:matrix}
\sum_{\rm spins}|{\cal M}|^2
=\left(\frac{C_A G_F}{\sqrt{2}}\right)^2 M_{\mu\nu} N^{\mu\nu},
\end{equation}
where $M^{\mu\nu}$ and $N^{\mu\nu}$ stand for the nucleonic and
neutrino parts, respectively.  From the discussion in Raffelt and
Seckel (1995) it follows that in an isotropic medium and for
nonrelativistic nucleons it is enough to consider the spatial trace
$\overline M\equiv\frac{1}{3}\,M_i^i$ which defines the ``reduced
squared matrix element.''  Equation~(\ref{eq:matrix}) may thus be
written as
\begin{equation}
\sum_{\rm spins}|{\cal M}|^2
\to \left(\frac{C_A G_F}{\sqrt{2}}\right)^2 
\overline M\,
8\omega_1\omega_2(3-\cos\theta),
\end{equation}
where $\omega_{1,2}$ are the neutrino energies and $\theta$ is the
angle between their momenta. 
With these simplifications and in the
long-wavelength limit the scattering kernel is
\begin{eqnarray}
S_\sigma^{(1)}(\omega)&=&
\frac{1}{n_B}
\int\prod_{i=1}^4 \frac{d^3{\bf p}_i}{2 m_N (2\pi)^3}
\,f_1 f_2 (1-f_3) (1-f_4)\,\overline M
\nonumber\\
&&\hskip3em
\times\,
(2\pi)^4\delta(E_1+E_2-E_3-E_4+\omega)
\,\delta^3({\bf p}_1+{\bf p}_2-{\bf p}_3-{\bf p}_4),
\label{kernel}
\end{eqnarray}
where $n_B$ is the baryon density. Further, $f_i$ is the occupation
number of nucleon $i$ with energy $E_i$ and momentum ${\bf p}_i$ while
$\omega=-(\omega_1+\omega_2)$ for pair emission and
$\omega=\omega_1+\omega_2$ for pair absorption, i.e.\ $\omega$ is the
transfer of energy to the nuclear medium.  For neutron-neutron or
proton-proton bremsstrahlung we need to include a statistics factor
$1/4$ to compensate for initial- and final-state double counting.
The superscript (1) signifies that in this form the scattering kernel
is a lowest-order perturbative result, derived from a bremsstrahlung
calculation.  The subscript $\sigma$ indicates that in the framework
of linear-response theory it is the nucleon spin-density
autocorrelation function (e.g.\ Janka, Keil, Raffelt \&\ Seckel 1996).
Both the lowest order $S^{(1)}_\sigma(\omega)$ and the full
spin-density autocorrelation function $S_\sigma(\omega)$ obey 
detailed balance 
$S_\sigma(-\omega)=S_\sigma(\omega)\,e^{-\omega/T}$.

In our treatment we always assume that the neutron spins evolve
independently of each other, i.e.\ we ignore possible spin-spin
correlations.  In this case the full scattering kernel must obey the
normalization requirement (Raffelt \& Strobel 1997)
\begin{equation}\label{norm}
\int_{-\infty}^{+\infty} \frac{d\omega}{2\pi}\,
S_\sigma(\omega)=\frac{1}{n_B}\int\frac{2d^3{\bf p}}{(2\pi)^3}\,
f_p(1-f_p) \equiv B
\end{equation}
where $f_p$ is the occupation number of a neutron with momentum ${\bf
p}$ and the factor 2 represents the two spin orientations.  In the
nondegenerate case where we may neglect the Pauli blocking factor
$(1-f_p)$ the normalization is unity.

\subsection{Nuclear Matrix Element}

\subsubsection{Identical Nucleons}

In order to calculate the scattering kernel we need to know the
nuclear matrix element.  The bremsstrahlung emission of neutrino pairs
or axions arises from nucleon spin fluctuations in collisions so that
one needs a spin-dependent nucleon-nucleon potential. The most general
velocity-independent interaction has a scalar (spin-independent) part,
a central part proportional to $\bs_1\cdot\bs_2$, and a tensor part
proportional to $3\hat{\bf r}\cdot\bs_1\,\hat{\bf
r}\cdot\bs_2-\bs_1\cdot\bs_2$ where $\bs_{1,2}$ are the spin operators
of the two nucleons (Blatt \& Weisskopf 1979, Ericson \& Weise 1988).
These interactions have the important property that they conserve the
square of the total nucleon spin, i.e.\ $(\bs_1+\bs_2)^2$ is a
constant of the motion.  In addition, the scalar and central parts
conserve the total spin $(\bs_1+\bs_2)$.

We first consider bremsstrahlung from the collision of identical
nucleons (proton-proton or neutron-neutron collisions).  The emission
of neutrino pairs or axions is induced by the total spin operator
$(\bs_1+\bs_2)$ which evolves nontrivially only due to the tensor
interaction. Because the physical cause of the tensor interaction is
primarily one-pion exchange (OPE) one expects that an OPE ansatz
captures the dominant aspect of the bremsstrahlung process (Friman \&
Maxwell 1979).  In addition, the spin operator has nonvanishing matrix
elements only between triplet states. Because the total wavefunction
must be antisymmetric the orbital angular momentum of the relevant
states must be odd. Therefore, $s$-waves do not contribute so that the
least well-known short-distance part of the nucleon interaction
potential does not affect the nucleon motion, further justifying the
OPE ansatz.

For identical nucleons the OPE squared matrix element in its
``reduced form'' is found to be (Raffelt \& Seckel 1995)
\begin{equation}\label{eq:opematrix}
\overline M
=\frac{64(4\pi)^2}{3}\,\frac{\alpha_\pi^2}{\omega^2}\,
\Biggl[\left(\frac{{\bf q}^2}{{\bf q}^2+m_\pi^2}\right)^2
+\left(\frac{{\bf q}_*^2}{{\bf q}_*^2+m_\pi^2}\right)^2
+\frac{{\bf q}^2{\bf q}_*^2-3({\bf q}\cdot{\bf q}_*)^2}
{({\bf q}^2+m_\pi^2)({\bf q}_*^2+m_\pi^2)}\Biggr],
\end{equation}
where $\omega=E_4+E_3-E_2-E_1$ is the energy transfer to the nucleons,
${\bf q}={\bf p_2}-{\bf p_4}$ is the momentum transfer between the
nucleons, and ${\bf q}_*={\bf p_2}-{\bf p_3}$ the momentum transfer
for the exchange amplitude.  Further, $\alpha_\pi\equiv
(f2m_N/m_\pi)^2/4\pi \approx15$ with $f\approx 1$ is the pion-nucleon
``fine-structure constant.''

At a temperature $T$ a typical nucleon momentum is $(3m_N T)^{1/2}$ so
that a typical momentum exchange in a collision is of a similar
magnitude. At $T=10\,\rm MeV$ this is about $170\,\rm MeV$, only
slightly larger than the pion mass of $135\,\rm MeV$. First, this
implies that the pion mass cannot be ignored in the denominators in
Eq.~(\ref{eq:opematrix}). Secondly, it means that the typical
potential region probed is not much smaller than $m_\pi^{-1}$ so that,
again, the OPE potential should be a reasonable
approximation. 

Two-pion exchange effects become important at distances below $2\,{\rm
fm} \simeq 1.5\,m_\pi^{-1}$ (Ericson \& Weise 1988).  Their impact on the
bremsstrahlung process can be estimated by mimicking the two-pion
exchange contribution by one-$\rho$-meson exchange where the mass of
this effective particle is taken to be $m_\rho\approx600\,\rm MeV$. A
typical term in Eq.~(\ref{eq:opematrix}) is then modified to
(Ericson \& Mathiot 1989)
\begin{equation}
\left(\frac{{\bf q}^2}{{\bf q}^2+m_\pi^2}\right)^2
\to\left(\frac{{\bf q}^2}{{\bf q}^2+m_\pi^2}
-C_\rho\frac{{\bf q}^2}{{\bf q}^2+m_\rho^2}\right)^2
\end{equation}
with $C_\rho=1.67$.  Taking ${\bf q}^2$ to be $3T m_N$ with $T=10\,\rm
MeV$, i.e.\ taking $|{\bf q}|$ to be around $170\,\rm MeV$ yields a
35\% reduction of the squared matrix element. This estimate quantifies
the error one is likely to make by using a simple OPE potential.

Even if the OPE potential is taken to be appropriate, this alone does
not guarantee that the Born approximation is justified which was used
to calculate Eq.~(\ref{eq:opematrix}). Ericson \& Weise (1988)
compare the $p$-wave scattering volumes of an OPE Born calculation
with an iterated OPE calculation and with a full calculation using the
Paris nucleon-nucleon potential. The OPE Born approximation, again,
yields a certain overestimate so that for our conditions of interest
the correct result may be smaller by as much as 50\%.
These uncertainties set the scale of precision that one may hope to
achieve with our simple approach.

The OPE approximation may be tested by comparing the calculated rate
for the pionic bremsstrahlung process $pp\to pp\pi^0$ with
experimental data (Choi, Kang \& Kim 1989; Turner, Kang \& Steigman
1989). The pion coupling is of derivative nature so that only the
axial-vector current contributes, rendering this process a good proxy
for the bremsstrahlung emission of neutrino pairs. However, because of
the pion mass threshold the kinematical regime probed is always at
higher (but not very much higher) energies than what is appropriate
for a SN core, and one needs to include relativistic corrections which
are no longer negligible. The agreement between an OPE calculation and
the data found by Turner, Kang \& Steigman (1989) is far better than
one would have expected according to our above caveats. After all, one
now probes the $NN$ interaction potential at distances where two-pion
exchange and other corrections surely must be important. This example
illustrates the well-known fact that the OPE potential in Born
approximation often yields far better results for spin-dependent
processes than one is entitled to expect.

\subsubsection{Proton-Neutron Scattering}

The situation is far more complicated for proton-neutron
scattering. Because the neutrino neutral-current coupling to protons
and neutrons is nearly equal but of opposite sign, the pair emission
is now induced essentially by the operator $\bs_1-\bs_2$ which is not
conserved by the central part of the interaction potential. This
operator connects triplet and singlet states, i.e.\ the selection
rules of the neutron-neutron system do not apply. Further, the
two-nucleon wave function can be antisymmetric in the isospin
variables so that orbital $s$-waves allow for both singlet and triplet
states. Put another way, the short-distance behavior of the potential
does matter in this context. By the same token, the comparison of
Turner, Kang \& Steigman (1989) with experimental data is not directly
relevant.

Recently Sigl (1997) studied numerically the behavior of
$S_\sigma(\omega)$ for this case, using a phenomenological potential
adapted to fit low-energy scattering data and deuteron properties. He
found huge differences between the exact $S_\sigma(\omega)$ and the
one derived in Born approximation. Put another way, because $s$-wave
scattering dominates in this problem, the phase shifts are large and
the Born approximation is not justified. For the chosen temperature
and density, processes of the form $p+n\to d+\nu+\bar\nu$ play a
dominant role while below the deuteron binding energy
($|\omega|<2.2\,\rm MeV$) only free-free transitions contribute to
$S_\sigma(\omega)$.  In this regime Sigl finds that his numerical
$S_\sigma(\omega)$ and the one derived from the OPE potential in Born
approximation agree surprisingly well---there must be compensating
effects between a potential and an approximation which are separately
unjustified.

Thus one may probably estimate the relative significance of the $pn$
process by using the OPE Born expression.  One finds for the squared
matrix element if one takes $|C_A|$ to be the same for protons and
neutrons (Raffelt \& Seckel 1995)
\begin{equation}
\Biggl[\left(\frac{{\bf q}^2}{{\bf q}^2+m_\pi^2}\right)^2
+2\left(\frac{{\bf q}_*^2}{{\bf q}_*^2+m_\pi^2}\right)^2
+2\,\frac{{\bf q}^2{\bf q}_*^2-({\bf q}\cdot{\bf q}_*)^2}
{({\bf q}^2+m_\pi^2)({\bf q}_*^2+m_\pi^2)}\Biggr]
\end{equation}
with all coefficients the same as in Eq.~(\ref{eq:opematrix}).  In the
limit of a vanishing pion mass the expression in square brackets of
Eq.~(\ref{eq:opematrix}) reduces to $3-\beta$ where $\beta$ is a
phase-space average of $3({\bf q}\cdot{\bf q}_*)^2/{\bf q}^2{\bf
q}_*^2$ which for nondegenerate neutrons is found to be
$\beta\approx1.31$. In the present case the expression in square
brackets reduces to $5-2\beta/3$. If we take for $\beta$ the same
value as before we find that the $pn$ squared matrix element is about
2.5 times as large as the $nn$ one.  For the $np$ process the
bremsstrahlung rate is proportional to the product of the densities
$n_p n_n$, i.e.\ to $n_B^2/4$ for an equal mix of protons and
neutrons. For pure neutrons it is also proportional
to $n_B^2/4$ where the factor $1/4$ now derives from the phase-space
reduction for identical particles. Put another way, the relative
importance of the two cases is indeed well estimated by a comparison
of the squared matrix elements.

For our conditions of interest protons are relatively rare so
that ignoring the $pn$ process leads to an underestimate of the
bremsstrahlung rate which is probably not larger than a few tens of
percent.  Therefore, the error made by considering a medium of
neutrons alone partially compensates the error made by using the OPE
potential in Born approximation. We are thus led to believe that the
compound error of our approximations does not exceed a few tens of
percent. 

\subsection{Generic Representation}

The squared matrix element Eq.~(\ref{eq:opematrix}) involves a factor
$\omega^{-2}$ which survives the nucleon phase-space integration and
which, in fact, is a generic feature of any bremsstrahlung process
(Raffelt 1996).  Therefore, we will always write the lowest-order
scattering kernel as
\begin{equation}\label{generic}
S_{\sigma}^{(1)}(\omega)=\frac{\Gamma_\sigma}{\omega^2}\,s(\omega/T)
\label{smalls}
\end{equation}
where $\Gamma_\sigma$ is what we call the ``spin fluctuation rate''
while $s(x)$ is a dimensionless, slowly varying function of order
unity.  The factorization between $\Gamma_\sigma$ and $s(x)$ is not
unique. We take $\Gamma_\sigma$ such that $s(0)=1$ when the neutrons
are nondegenerate and when the pion mass has been ignored in the
matrix element, leading to (Raffelt and Seckel 1995)
\begin{equation}\label{eq:gamsig}
\Gamma_\sigma
=\frac{8\sqrt{2\pi}\,\alpha_\pi^2}{3\pi^2}\,\eta_*^{3/2}\,
\frac{T^2}{m_N}.
\end{equation}
In terms of the neutron Fermi momentum $p_{F}$ the ``effective
degeneracy parameter'' is
\begin{equation}\label{eq:effdeg}
\eta_{\ast} \equiv \frac{p_{F}^2}{2 m_N T};
\end{equation}
it varies relatively little throughout a SN core and thus is a
convenient measure of the neutron density.  All modifications of
$S_\sigma^{(1)}(\omega)$ by neutron degeneracy and the finite pion
mass are included in the dimensionless function $s(x)$ which will be
determined in Sec.~4. 

For bremsstrahlung processes the singular behavior of
$S^{(1)}_\sigma(\omega)$ at $\omega=0$ is of no concern because it is
suppressed by a sufficiently high power of $\omega$ in all relevant
neutrino phase-space integrations. However, we want a kernel which
consistently describes bremsstrahlung {\it and\/} the ``inelasticity''
of scattering. Even in this case the singularity can be interpreted
under a phase-space integral such that one obtains finite and
meaningful results (Sawyer 1995; Raffelt, Seckel \& Sigl 1996). Still,
the singular behavior is an artifact of the lowest-order perturbative
expansion. Multiple-scattering effects (or formally a resummation of
the neutron propagator) render $S_\sigma(\omega)$ a well-behaved
function everywhere.

The interpretation of $S_\sigma(\omega)$ as a spin
autocorrelation function suggests a phenomenological inclusion of
multiple-scattering effects by the ansatz
\begin{equation}\label{eq:lorentzian}
S_\sigma(\omega)=\frac{\Gamma_\sigma}{\omega^2+\Gamma^2/4}\,
s(\omega/T).
\end{equation}
In the classical limit of ``hard collisions'' one has $s(x)=1$ and
$\Gamma=\Gamma_\sigma$, leading to the correct normalization.  In the
general case we choose $\Gamma$ such that $S_\sigma(\omega)$ is
properly normalized.

To summarize, our scattering kernel is determined by a two-step
procedure. We first calculate the perturbative
$S^{(1)}_\sigma(\omega)$ by what amounts to a bremsstrahlung
calculation and from which we extract $s(x)$.  Next, we determine
$\Gamma$ in Eq.~(\ref{eq:lorentzian}) such that the normalization
Eq.~(\ref{norm}) is fulfilled. We stress that the exact shape of
$S_\sigma(\omega)$ around $\omega=0$ is not crucial; in that sense our
Lorentzian ansatz is well-motivated but arbitrary. The important
quality of our ansatz is that it gives the correct normalization and
thus the correct neutrino scattering cross section. 

\subsection{Neutrino Processes}

\subsubsection{Boltzmann Collision Integral}

We have defined the scattering kernel essentially as the squared
matrix element of a neutrino-nucleon process with the nucleon degrees
of freedom integrated out. In order to calculate quantities like a
neutrino mean free path it remains to integrate over the neutrino
phase space. To this end we begin with the
Boltzmann collision integral for neutrinos
\begin{eqnarray}\label{eq:boltzmann}
\dot f_1\Big|_{\rm coll}
&=&C_A^2 G_F^2n_B
\int\frac{d^3 {\bf k}_2}{(2\pi)^3}\,(3-\cos\theta)\nonumber\\
&&\hskip4em\times\,
\Bigl[(1-f_1)(1-\bar f_2)\,S_\sigma(-\omega_1-\omega_2)
-\,f_1\bar f_2 S_\sigma(\omega_1+\omega_2)\nonumber\\
&&\hskip5em+\,
(1-f_1)f_2\,S_\sigma(-\omega_1+\omega_2)
-f_1(1-f_2)\,S_\sigma(\omega_1-\omega_2)\Bigr],
\end{eqnarray}
where $f_{1,2}$ are now the occupation numbers of neutrinos with
momenta ${\bf k}_1$ and ${\bf k}_2$, respectively, while $\bar
f_{1,2}$ are those of antineutrinos with the corresponding
momenta. The first two terms in square brackets represent pair
emission and pair absorption by the medium, i.e.\ the bremsstrahlung
process and its inverse, while the third and fourth term represent the
gain and loss terms from neutrino scattering on the nuclear medium.
An analogous equation obtains for antineutrinos, i.e.\ for $\bar f_1$.

For a practical numerical implementation we mention that the dimension
of $S_\sigma$ is (energy)$^{-1}$ so that an integral $\int
d\omega\,S_\sigma(\omega)$ is a dimensionless number. Further, in
natural units we may write $d^3{\bf k}_2= d\Omega_2\omega_2^2
d\omega_2$ so that in the units usually employed in SN physics the
integral expression in Eq.~(\ref{eq:boltzmann}) has the dimension $\rm
MeV^2$. In the overall coefficient we write the baryon density as
$n_B=\rho/m_N$ in terms of the mass density and the vacuum nucleon
mass. Then the overall coefficient is numerically $C_A^2G_F^2 n_B
=3.79\times 10^4\,{\rm s}^{-1}\,{\rm MeV}^{-2} \,\rho_{14}$, where
$\rho_{14}=\rho/10^{14}\,{\rm g\,cm^{-3}}$ and where we have used
$C_A=-1.26/2$.

\subsubsection{Bremsstrahlung}

To illustrate the use of the scattering kernel and for later reference
we turn to a few simple quantities related to bremsstrahlung.  The
inverse mean free path of a neutrino with energy $\omega_1$ against
pair-absorption is
\begin{equation}
\lambda_1^{-1}=
C_A^2 G_F^2 n_B
\int\frac{d^3 {\bf k}_2}{(2\pi)^3}\,(3-\cos\theta)\,\bar f_2
S_\sigma(\omega_1+\omega_2)
\end{equation}
with $\bar f_2$ the occupation number of the antineutrinos.  We use a
Maxwell-Boltzmann distribution at temperature $T$ for the
neutrinos. With the representation Eq.~(\ref{generic}) for the
scattering kernel the spectral average is
\begin{equation}\label{simple}
\Gamma_{\nu_\mu\bar\nu_\mu nn\to nn}=
\langle \lambda_1^{-1} \rangle = 
\frac{3C_A^2 G_F^2}{\pi}\,n_B T^2\,\frac{\Gamma_\sigma/T}{20\pi}\,\xi
\end{equation}
with the ``dimensionless mean free path''
\begin{equation}
\xi\equiv 5
\int_{0}^{\infty}dx_1 \int_{0}^{\infty}dx_2\,e^{-x_1}e^{-x_2}\, 
\frac{x_1^2 x_2^2}{(x_1+x_2)^2}\,s(x_1+x_2).
\label{xi}
\end{equation}
Here, $x_1 \equiv \omega_1/T$ and $x_2 \equiv \omega_2/T$ are the
dimensionless neutrino and antineutrino energies respectively and we
have ignored multiple-scattering effects. One of the energy
integrations can be done analytically so that
\begin{equation}\label{xidef}
\xi=\frac{1}{6}
\int_{0}^{\infty}dx\,x^3 e^{-x}\,s(x).
\end{equation}
Taking $s(x)=1$ leads to $\xi=1$; for realistic conditions it is
in the range 0.2--0.5 (Sec.~4.5).

Next we consider the energy-loss rate of a medium which is taken to be
transparent to neutrinos so that one may ignore phase-space blocking
effects,
\begin{eqnarray}
Q_{nn\to nn\nu\bar\nu}&=&
C_A^2 G_F^2 n_B
\int\frac{d^3 {\bf k}_1}{(2\pi)^3}
\frac{d^3 {\bf k}_2}{(2\pi)^3}\,(3-\cos\theta)\,
(\omega_1+\omega_2)\,
S_\sigma(-\omega_1-\omega_2)
\nonumber\\
&=&\frac{C_A^2 G_F^2 n_B}{40\pi^4}
\int_0^\infty d\omega\,\omega^6\,e^{-\omega/T}
S_\sigma(\omega).
\label{eq:emission}
\end{eqnarray}
The production rate of neutrino pairs instead of the energy-loss rate
is found by dropping the energy $\omega=(\omega_1+\omega_2)$ in this
expression. 

In numerical SN codes it is sometimes useful to consider the
differential energy production rate into the neutrino channel only,
not counting the energy which goes into antineutrinos. Therefore,
we must integrate over the $\bar\nu$ phase-space, ignoring for
simplicity Pauli blocking effects,
\begin{equation}
\frac{dQ_{\rm brems}}{d\omega_1}
=\frac{3 C_A^2 G_F^2 n_B}{4 \pi^4} \,
\omega_1^3\int_{\omega_1}^\infty d\omega\,(\omega-\omega_1)^2
e^{-\omega/T}\,S_\sigma(\omega)\,.
\end{equation}
The integral over this quantity is half of $Q_{nn\to nn\nu
\bar{\nu}}$ of Eq.~(\ref{eq:emission}) because we now measure only the
energy carried away by the neutrino, ignoring that carried by the
antineutrino.  In proper units we have $C_A^2 G_F^2
n_B=1.13\times10^{31}\,{\rm ergs\,cm^{-3}\,s^{-1}
\,MeV^{-6}}\,\rho_{14}$.

In order to compare the efficiency of energy transfer of different
processes it will turn out to be useful to consider the net transfer
of energy between a neutron fluid at temperature $T$ to a neutrino
fluid at temperature $T_\nu$. It is found to be
\begin{equation}
\Delta Q_{nn\leftrightarrow nn\nu\bar\nu}=
\frac{C_A^2 G_F^2 n_B}{40\pi^4}
\int_0^\infty d\omega\,\omega^6\,
\left(e^{-\omega/T}-e^{-\omega/T_\nu}\right)
S_\sigma(\omega).
\end{equation}
With $\Delta T=T-T_\nu$ and in the limit $|\Delta T|\ll T$ this is
\begin{eqnarray}\label{bremstransfer}
\frac{\Delta Q_{nn\leftrightarrow nn\nu\bar\nu}}{\Delta T}
&=&\frac{C_A^2 G_F^2 n_B}{40\pi^4}\,\frac{1}{T^2}
\int_0^\infty d\omega\,\omega^7\,e^{-\omega/T}\,S_\sigma(\omega)
\nonumber\\
&\simeq&\frac{3C_A^2 G_F^2 n_B T^4}{\pi^4}\,\Gamma_\sigma
\int_0^\infty dx\,\frac{x^5\,e^{-x}}{120}\,s(x),
\end{eqnarray}
where we have used the representation Eq.~(\ref{generic}) for the
lowest-order scattering kernel which may be used because the high
power of $\omega$ appearing under the integral renders the
low-$\omega$ modification by multiple-scattering irrelevant.  With
$s(x)=1$ the integral is~1.

\subsubsection{Neutrino Scattering}

When studying the bremsstrahlung process one is naturally led to the
Boltzmann collision integral of Eq.~(\ref{eq:boltzmann}) which would
be incomplete without the scattering processes.  For a neutrino of
energy $\omega_1$ the scattering cross section differential with
regard to the final-state energy $\omega_2$ and the scattering angle
is easily identified to be
\begin{equation}
\frac{d\sigma_1}{d\omega_2 d\Omega_2}=
\frac{3-\cos\theta}{4\pi}
\frac{C_A^2 G_F^2}{\pi}
\,\omega_2^2\,\frac{S_\sigma(\omega_1-\omega_2)}{2\pi}.
\end{equation}
Numerically we have $C_A^2 G_F^2 = 2.10\times10^{-44}\,\rm cm^2\,
MeV^{-2}$.  If the medium is dilute so that the spin-fluctuation rate
is small, then $S_\sigma(\omega)/2\pi$ is strongly peaked around
$\omega=0$ and actually must approach a $\delta$-function which is
normalized according to Eq.~(\ref{norm}). Integrating over energy and
angles then gives us the usual elastic scattering cross section
$\sigma_1= (3 C_A^2 G_F^2/\pi) \omega_1^2$ which in a degenerate
medium is reduced according to the right-hand side of
Eq.~(\ref{norm}).

Next we calculate the total axial-current cross section, averaged over
a Maxwell-Boltzmann neutrino distribution. We find, in agreement with
Raffelt \& Seckel (1995),
\begin{equation}\label{cross-section}
\langle \sigma_A\rangle
=\frac{3 C_A^2 G_F^2}{\pi}\,12 T^2\,
\int_0^\infty d\omega \,
\frac{12+6\omega/T+\omega^2/T^2}{12\pi}\,
e^{-\omega/T}\,S_\sigma(\omega).
\end{equation}
In the dilute limit where $S_\sigma(\omega)\to 2\pi \delta(\omega)$
the integral is~1. In general it is less than 1 because the function
$S_\sigma(\omega)$ is normalized to $1$ while the rest of the
integrand falls monotonically with $\omega$ so that the integral has
its maximum value when $S_\sigma$ is narrowly peaked around
$\omega=0$. Therefore, $NN$ interactions lead to a reduction of the
average neutrino scattering cross section. 
This cross-section reduction is an unavoidable consequence of $NN$
interactions and occurs on the same level of approximation as the
bremsstrahlung emission or absorption of neutrino pairs.  

In order to compare the efficiency of energy transfer of the
``inelastic scattering process'' with that of bremsstrahlung and other
processes we finally consider the rate of energy transfer between a
nucleon and a neutrino fluid at slightly different temperatures.  We
find
\begin{equation}\label{scatteringtransfer}
\frac{\Delta Q_{\nu nn\leftrightarrow nn\nu}}{\Delta T}
=\frac{30\,C_A^2 G_F^2 n_B T^4}{\pi^4}\,\Gamma_\sigma
\int_0^\infty dx\,\frac{12+6x+x^2}{20}\,e^{-x}\,s(x),
\end{equation}
where we have used Eq.~(\ref{generic}) for the lowest-order scattering
kernel. With $s(x)=1$ the integral is~1.

\section{Analytic Fitting Formula}

\subsection{General Representation}

In order to represent the scattering kernel by analytic fitting
formulae we write it in the form
\begin{equation}\label{eq:generic}
S_\sigma(\omega)= \frac{1}{T}\,
\frac{\gamma}{x^2+(\gamma g/2)^2}\,s(x)
\end{equation}
where $x=\omega/T$. The dimensionless quantities $\gamma$, $g$, and
$s(x)$ depend on the medium density and temperature. We characterize
the former by the neutron effective degeneracy parameter introduced in
Eq.~(\ref{eq:effdeg}) which may be written as 
\begin{equation}
\eta_*=\frac{(3\pi^2)^{2/3}}{2 m_N T}\,
\left(\frac{\rho}{m_N}\right)^{2/3}
=3.04\,\rho_{14}^{2/3}\,T_{10}^{-1}.
\end{equation}
The spin-fluctuation rate of Eq.~(\ref{eq:gamsig}) is numerically
\begin{equation}\label{numericalgamma}
\gamma\equiv\frac{\Gamma_\sigma}{T}
=1.63\,\eta_*^{3/2} T_{10}
=8.6\,\rho_{14}\,T_{10}^{-1/2}.
\end{equation}
Both $g$ and $s(x)$ are always of order unity so that simple estimates
of bremsstrahlung-related quantities may be obtained by setting them
both equal to 1.

The matrix element Eq.~(\ref{eq:opematrix}) involves the pion mass as
one more dimensionful parameter. It proves useful to express it in
the form
\begin{equation}
y\equiv \frac{m_\pi^2}{m_N T}=1.94\,T_{10}^{-1}.
\end{equation}
With this notation we have that the dimensionless scattering kernel
$s(x)$ and the quantity $g$ depend on the parameters $y$ and $\eta_*$.

\subsection{Nondegenerate Limit}

We will construct a general analytic representation from an
interpolation between the nondegenerate (ND) and degenerate (D)
limiting cases. The former was treated by Raffelt \& Seckel (1995);
in a symmetric form their result can be expressed as
\begin{eqnarray}\label{lastintegral}
&&\hskip-2em \bar s_{\rm ND}(x,y)\equiv
 \frac{s_{\rm ND}(x,y)+s_{\rm ND}(-x,y)}{2}={}\nonumber\\
\noalign{\vskip4pt}
&&\hskip-2em\frac{e^{-x/2}+e^{x/2}}{16}
\int_{|x|}^{\infty}
dt\,\,e^{-t/2}\,\,\frac{3x^2+6ty+5y^2}{3}
\Biggl[\frac{2\sqrt{t^2-x^2}}{x^2+2ty+y^2}
-\frac{1}{t+y}\,
\log\left(\frac{t+y+\sqrt{t^2-x^2}}{t+y-\sqrt{t^2-x^2}}\right)\Biggr].
\nonumber\\
\end{eqnarray}
Evidently, this expression is even in $x$; the structure function
is recovered by $s_{\rm ND}(x,y)=2\bar s_{\rm ND}(x,y)/(1+e^{-x})$
and then obeys detailed balance. Its limiting behavior is
\begin{equation}
s_{\rm ND}(x,y)=\cases{1&for $x=0$ and $y=0$,\cr
  (4\pi/x)^{1/2}&for $x\gg1$ and $x\gg y$,\cr
  \frac{80}{3}\, y^{-2}&for $x=0$ and $y\gg1$,\cr}
\end{equation}
i.e.\ it falls off to zero for either large $x$ or $y$.

In order to construct an analytical fit we first extract the
detailed-balance behavior explicitly by
\begin{equation}
s(x,y) = \hat{s}(x,y) 
\times \cases{1 & for $x > 0$,\cr
e^{-x} & for $x < 0$. \cr}
\end{equation}
An analytical fit for the $y=0$ case (vanishing pion mass) is
\begin{equation}\label{ndkernel}
\hat s_{y=0}(x)\simeq
\left(\frac{x}{4\pi}+\left[1+\left(12+\frac{3}{\pi}\right)x
\right]^{-1/12}\right)^{-1/2}.
\end{equation}
This function reproduces the limiting behavior at $x=0$ and $x\gg1$,
it has the correct derivative $1/2$ at $x=0$, and it deviates from the
correct result by no more than 2.5\% anywhere.  Likewise we have
constructed
\begin{equation}
\hat s_{x=0}(y)\simeq
\Biggl[1+\frac{y}{2}\log\left(\frac{y}{1+y}\right)\Biggr]
\Biggl[1+\left(\frac{3 y^2}{160}\right)^{4/9}\Biggr]^{-9/4}.
\end{equation}
It has the correct limiting behavior at $y=0$ and $y\gg1$, it is
vertical at $y=0$ in agreement with the full result, and its maximum
error is below 4.5\% anywhere.  This function gives us a good idea of
the suppression caused by including the pion mass.

A simple estimate of the compound scattering kernel is $\hat s_{\rm
ND}(x,y)\simeq\hat s_{y=0}(x)\, \hat s_{x=0}(y)$. However, it
underestimates the correct value by as much as a factor of a few when
both $x$ and $y$ are a few which is the most relevant regime in a SN
core.  To construct a more suitable approximation we expand the square
bracket in Eq.~(\ref{lastintegral}) in a power series in $t$ and keep
only the lowest term which is proportional to $t^{3/2}$. Now the
exponential can be integrated so that
\begin{equation}
\hat s_{\rm ND}(x,y)\simeq
\frac{2\pi^{1/2} x^{3/2}(3 x^2+6xy+5y^2)}{3(x+y)^4},
\end{equation}
valid for $x\gg 1$ and $y\gg 1$.  With a little bit of tinkering one
can supplement it such that it behaves reasonably for small $x$ and
$y$,
\begin{equation}
\hat s_{\rm ND}(x,y)\simeq
\frac{2\pi^{1/2} \left(x+2-e^{-y/12}\right)^{3/2}
\left(x^2+2xy+\frac{5}{3}y^2 + 1\right)}
{\pi^{1/2}+(\pi^{1/8}+x+y)^4}.
\end{equation}
This function is 1 for $x=y=0$ and has the correct limiting behavior
for $y=0$ and $x\gg 1$, but it does not have the correct limiting
behavior for $x=0$ and $y\gg 1$. The maximum error for any $x$ is
14\%, 7\%, 5\%, 9\%, and 12\% for $y=0.5$, 1, 2, 3, and 5,
respectively. The errors are even smaller for an integrated quantity
like the average mean free path of Eq.~(\ref{xidef}) where we find a
deviation of $-4\%$, $-1\%$, 4\%, 7\% and 10\% for the same $y$
values. This precision is good enough in view of the relatively crude
treatment of the nuclear matrix element.

\subsection{Degenerate Limit}

In the degenerate limit the phase space integration can be carried out
analytically even with a nonzero pion mass. 
Ishizuka \& Yoshimura (1990) found
\begin{equation}
s_{\rm D}(x,y,\eta_{\ast}) = 
3 \left(\frac{\pi}{2}\right)^{5/2} \eta_{\ast}^{-5/2}\, 
\frac{(x^2+4 \pi^2)x}{4 \pi^2(1-e^{-x})}\, 
f\left(\sqrt{\frac{y}{2\eta_{\ast}}}\right),
\end{equation}
where
\begin{equation}
f(u) = 1-\frac{5u}{6}\arctan \left(\frac{2}{u}\right)+
\frac{u^2}{3(u^2+4)}+\frac{u^2}{6\sqrt{2u^2+4}}\arctan
\left(\frac{2\sqrt{2u^2+4}}{u^2}\right).
\end{equation}
The function $s_{\rm D}(x)$ fulfills detailed balance explicitly.
However, it is pathological in that it scales as $x^3$ for $x\gg 1$
which implies that the full scattering kernel is not normalizable in
the sense of Eq.~(\ref{norm}). Of course, the assumption of degenerate
nucleons is never good for energy transfers which far exceed the
nucleon Fermi energy. For $\omega\gg E_F$ ($x\gg E_F/T$) the full
$s(x)$ must always approach $s_{\rm ND}(x)$.

\subsection{Interpolation}

An interpolation which is accurate to within roughly 30--40\% for
all values of $x$ and $y$ is provided by
\begin{equation}
s=\left(\frac{1}{s_{\rm ND}} + \frac{1}{s_{\rm D}}\right)^{-1}.
\label{simpleint}
\end{equation}
One can do better by using
\begin{equation}
s=\left(s_{\rm ND}^{-p(y)} +s_{\rm D}^{-p(y)}\right)^{-1/{p(y)}}
F(x,y,\eta_{\ast})\,
\Bigl[1+ C(x,y,\eta_*)\,G(x,y,\eta_{\ast})\Bigr],
\label{analyt}
\end{equation}
where
\begin{eqnarray}
F(x,y,\eta_{\ast}) & = & 1 + \frac{1}{[3+(x-1.2)^2+x^{-4}]\,
(1+\eta_{\ast}^2)\,(1+y^4)},\nonumber \\
G(x,y,\eta_{\ast}) & = & 
1-0.0044\,x^{1.1}\,\frac{y}{0.8+0.06\,y^{1.05}}\,
\frac{\eta_*^{0.5}}{\eta_*+0.2},
\nonumber\\ 
C(x,y,\eta_*)&=&\frac{1.1 \, x^{1.1}\,h(\eta_{\ast})}
{2.3 + h(\eta_{\ast}) x^{0.93} + 0.0001\, x^{1.2}}\,
\frac{30}{30+0.005\,x^{2.8}},\nonumber\\
p(y) & = & 0.67 + 0.18 \, y^{0.4},\nonumber\\
h(\eta_{\ast})&=&\frac{0.1\, \eta_{\ast}}{2.39+0.1\, 
\eta_{\ast}^{1.1}}.
\end{eqnarray}
This interpolation function reproduces the true value of the
scattering kernel to within 5--10\% in the physically interesting
parameter space, $y \simeq 2$--6 and $\eta_{\ast} \simeq 2$--6, and
maximum deviations for any value of the parameters of less than 25\%.
As discussed in Sec.~3.3, the uncertainties from the nuclear matrix
element are at least of that magnitude so that this fitting accuracy
is entirely sufficient.

In Table 2 we give $s(x,y,\eta_*)$, both the numerically
calculated values and for comparison the analytical fit.  In Fig.~3 we
plot $s(x)$ for different values of $y$ and $\eta_{\ast}$ where the
solid lines represent the numerical result and the short-dashed ones
the fitting formula. The detailed-balance condition tells us that
$s(-x)=s(x) e^{-x}$. Still, we plot $s(x)$ somewhat across $x=0$ in
order to illustrate that it is smooth at the origin.  In both panels
the upper curve corresponds to a vanishing pion mass ($y=0$) and
essentially nondegenerate conditions ($\eta_*=0.31$). 

The upper panel illustrates the effect of degeneracy which suppresses
the scattering kernel at low $x$, but enhances it at large ones.  All
of these curves approach the nondegenerate case asymptotically in the
limit $x\to\infty$, but they do so from above.  The lower panel
illustrates the impact of including the pion mass in the matrix
element. It suppresses the scattering kernel at low $x$ by an amount
which depends on $y$, and again the curves approach the $y=0$ case in
the large $x$ limit, however here they do so from below.

\subsection{Multiple Scattering}

It remains to determine the function $g(y,\eta_*)$ for the denominator
of Eq.~(\ref{eq:generic}) which must be adjusted to meet the
normalization condition Eq.~(\ref{norm}),
\begin{equation}\label{eq:normalization}
\int_{-\infty}^{+\infty} \frac{dx}{2\pi}\,
\frac{\gamma}{x^2+(\gamma g/2)^2}\,s(x) = B.
\end{equation}
For ND conditions and when $\gamma\ll 1$ the Lorentzian is essentially
a $\delta$ function and we have for the integral $s_{\rm ND}(0)/g$.
Therefore, in the limit $y=0$ and $\eta_*=0$, which also implies
$\gamma=0$, we have $g=1$ because $\gamma$ was defined such that in
this limit $s_{\rm ND}(0)=1$.

We have determined the function $g(y,\eta_{\ast})$ using the
normalization condition Eq.\ (\ref{norm}) and our analytic
approximation formula Eq.\ (\ref{analyt}) for $s(x)$.  Furthermore we
need to provide an analytic approximation for $g(y,\eta_{\ast})$ that
can be implemented in numerical calculations.  For very high values of
$y$, we know that $g \propto y^{-2}$ and for $y \to 0$ we know that $g
\to 0.5$ in the degenerate limit and $g \to 1$ in the nondegenerate
limit.  A decent analytic fit that has the correct limiting behavior
is given by
\begin{equation}
g(y,\eta_{\ast}) = \frac{\alpha_1 + \alpha_2 y^{p_1}}
{1+\alpha_3 y^{p_2} + \alpha_2 y^{p_1+2}/13.75},
\label{multia}
\end{equation}
where the coefficients are given by
\begin{eqnarray}
\alpha_1 & = & \frac{0.5+\eta^{-1}}{1+\eta^{-1}} \frac{1}{25 \, y^2 + 1}
+ (0.5 + \eta/15.6) \frac{25 \, y^2}{25 \, y^2 + 1},\nonumber \\
\alpha_2 & = & \frac{0.63 + 0.04 \, \eta^{1.45}}{1+0.02 \,
\eta^{2.5}},
\nonumber\\
\alpha_3 & = & 1.2 \, \exp(0.6 \, \eta - 0.4 \, \eta^{1.5}),\nonumber\\
p_1 & = & \frac{1.8 + 0.45 \, \eta}{1+0.15 \, \eta^{1.5}},\nonumber\\
p_2 & = & 2.3 - \frac{0.05 \, \eta}{1+0.025 \, \eta}.
\end{eqnarray}
In Fig.~4 we show $g(y,\eta_{\ast})$ as a function of $y$ for
different values of the degeneracy parameter $\eta_{\ast}$ (solid
lines) as well as our analytic fit to $g(y,\eta_{\ast})$ (dashes).
Also, in Fig.~5 we show the normalization $B(\eta_*)$ which was 
defined in Eq.~(\ref{norm}) as a function of $\eta_*$.

We may now calculate the average neutrino absorption rate in its
dimensionless form Eq.~(\ref{xidef}).  In Table~3 we give values for
$\xi$ based on our analytic approximation scheme, both with and
without the inclusion of multiple-scattering effects.  For realistic
SN conditions where $y\simeq2$--6 and $\eta_{\ast}\simeq 2$--6
we see that $\xi \simeq 0.2$--0.5. Multiple scattering is not a
strong effect for the average absorption rate.

However, to calculate inelastic scattering it is unavoidable to
include the multiple scattering effect. With our function 
$g(y,\eta_*)$ the scattering kernel is self-consistent in that it
allows for the simultaneous treatment of bremsstrahlung and inelastic
scattering while reproducing the correct total scattering cross
section. 


\section{Conclusions}

We have studied the neutron-neutron bremsstrahlung process
$nn\leftrightarrow nn\nu\bar\nu$ as an opacity source for neutrinos
and its impact on the neutrino spectra formation in supernovae.  For
$\nu_{\mu}$ and $\nu_{\tau}$, bremsstrahlung is by far the most
important number-changing reaction, far more important than $e^+e^-$
annihilation or the plasma process. As an energy-changing process it
is roughly as strong as elastic scattering on electrons and positrons.
We estimate that including bremsstrahlung will reduce the $\nu_\mu$
temperature by more than $1\,\rm MeV$. This is only a differential
effect in the sense that the $\nu_\mu$ and $\bar\nu_e$ spectral
temperatures become closer by this amount. Their absolute changes can
not be determined by our simple procedure.  The overall magnitude of
our differential effect is in agreement with Suzuki's (1991, 1993)
numerical simulations.

If one includes the bremsstrahlung process it is inconsistent to
ignore the inelasticity of $\nu n$ scattering which may be pictured as
the ``crossed bremsstrahlung process'' $\nu nn\to nn \nu$. This
process cannot change the neutrino number density, but as a source of
spectral equilibration it is roughly a factor of 10 more important
than bremsstrahlung and thus also far more important than $\nu e$
scattering. Moreover, depending on details of the temperature and
density profile of the SN core surface layers, the nucleon recoil in
$\nu N$ collisions is of roughly equal importance to the inelasticity
(Janka et~al.~1996).  Bremsstrahlung, inelastic scattering, and
recoils near the neutrino spheres is likely to make the spectra and
fluxes of the different neutrino species far more similar than had
been thought previously. This would modify all phenomena related to
neutrino flavor oscillations which rely on the difference in the
spectral temperatures to be effective.

The main purpose of our paper was to provide an explicit form for the
scattering kernel $S_\sigma(\omega)$ that will allow one to study the
impact of bremsstrahlung and inelastic scattering consistently in a
numerical simulation.

Our derivation is fundamentally perturbative; it is useful to study
the spectra formation in the surface layers of a proto neutron star.
We believe our scattering kernel to be appropriate for densities
roughly below $10^{14}\,\rm g\,cm^{-3}$.  At higher densities our
result predicts a huge suppression of the total $\nu N$ scattering
cross section which opens the thorny issue of its actual
magnitude. This problem has not been solved so that any SN simulation
has to rely on an arbitrary prescription for the neutrino diffusion
coefficients in the deep interior.  The SN 1987A signal (Keil, Janka
\& Raffelt 1995) as well as the $f$-sum rule for the spin-density
structure function (Sigl 1996) indicate that the cross-section
suppression is not as large as predicted by our scattering kernel, but
how large the cross sections truly are is not known. Recent attempts
to include the correct nucleon dispersion relation as well as hyperons
(Reddy \& Prakash 1997) predict modifications of the ``standard cross
sections'' by factors of order unity, but no attempt was made to
address the modifications caused by the inevitable nucleon or hyperon
spin-spin interactions.

Another shortcoming of our scattering kernel is that it does not
include recoil effects. In some regions of a SN core it may be even
more effective than $NN$ interactions at modifying the neutrino energy
in a collision.  A scattering kernel which includes recoils and $NN$
interactions simultaneously requires avoiding the long-wavelength
approximation, a rather complicated task that would imply another
variable (the scattering angle) explicitly in the scattering kernel.

Our estimates clearly show that bremsstrahlung and its related
processes cannot be ignored for the formation of the neutrino spectra
in a SN core.  However, to determine the real magnitude of the
modification one needs to perform a self-consistent numerical analysis
which includes feedback effects on the medium.  If such a study
reveals that the modifications are as severe as suggested by our
estimates one may be motivated to derive a more complete perturbative
scattering kernel. One could study the proton-neutron process on the
basis of a realistic nucleon-nucleon potential in the spirit of Sigl's
(1997) recent work, thereby avoiding the pathologies inherent in an
OPE Born treatment of $np$ scattering.  Moreover, it would be
cumbersome but straightforward to include recoil effects and $NN$
interactions simultaneously in the scattering kernel.  Unfortunately,
it remains far from obvious how to proceed extending the scattering
kernel into the high-density regime.


\section*{Acknowledgements}

We thank Thomas Janka and Adam Burrows for invaluable comments on
early versions of the manuscript and Wolfram Weise for discussions
about the use of the one-pion exchange potential.  We are especially
indebted to David Seckel for prodding us to include a discussion of
inelastic scattering and recoils.  Our research was supported, in
part, by the European Union under contract No.\ CHRX CT930120, by the
Deutsche Forschungsgemeinschaft under grant No.\ SFB~375, and by the
Theoretical Astrophysics Center under the Danish National Research
Council.


\appendix

\section{Nucleon Phase-Space Integration}

We give some details of solving the nucleon phase-space integral
Eq.~(\ref{kernel}).  From the original 12-dimensional integral one can
integrate out 8 dimensions analytically so that in the end
only a four dimensional integral remains to be done numerically.
First, the three-dimensional momentum delta function is used to
integrate out $d^3{\bf p}_4$. Because the nucleons are 
nonrelativistic their energy is $E_i = p_i^2/2m_N$ so that
\begin{equation}
E_1+E_2-E_3-E_4+\omega = 
\frac{- 2 p_3^2 - 2 {\bf p}_1 \cdot {\bf p}_2 +
2 {\bf p}_1 \cdot {\bf p}_3 + 2 {\bf p}_2 \cdot {\bf p}_3 }
{2 m_N} + \omega.
\end{equation}
Because of the assumed isotropy of the medium the ${\bf p}_1$ momentum
may be chosen in the $z$-direction, i.e.\ we have trivially $\int
d^3{\bf p_1}=4\pi \int dp_1$ with $p_1=|{\bf p}_1|$. Next we use polar
coordinates with $\alpha$ and $\beta$ the polar and azimuthal angles
of ${\bf p}_2$ relative to ${\bf p}_1$ and $\theta$ and $\phi$ those
of ${\bf p}_3$ so that
\begin{eqnarray}
d^{3}{\bf p}_{2} = p_{2}^{2}\,dp_{2}\,d\cos\alpha\, d\beta \\ 
d^{3}{\bf p}_{3} = p_{3}^{2}\,dp_{3}\,d\cos\theta\, d\phi.
\end{eqnarray}
Because of the medium's isotropy one of the azimuthal integrations is
trivial. We chose the $d\phi$ integration so that three nontrivial
angular integrations remain. In terms of these remaining angular
variables we have
\begin{eqnarray}
{\bf p}_1 \cdot {\bf p}_2 & = & p_1 p_2 \cos \alpha \\
{\bf p}_1 \cdot {\bf p}_3 & = & p_1 p_3 \cos \theta \\
{\bf p}_2 \cdot {\bf p}_3 & = & p_2 p_3 \cos \alpha \cos \theta + 
\sin \alpha \sin \theta \cos \beta.
\end{eqnarray}

The integration over $d \beta$ is carried out using the 
$\delta$-function where we write
$f(\beta)\equiv E_1+E_2-E_3-E_4+\omega$
so that
\begin{equation}
\label{beta}
\int_0^{2\pi} d \beta\, \delta (f(\beta)) = 
\sum_i \int_0^{2\pi} d \beta\, \frac{1}
{\left| \frac{df(\beta)}{d \beta} \right| _{\beta = \beta_{i}}}
\,\delta (\beta - \beta_{i}),
\label{betaint}
\end{equation}
where $\beta_{i}$ with $i=1,2$ are the two roots of $f(\beta)=0$, one
in the interval $[0,\pi]$ and one in the interval $[\pi,2\pi]$.
Because Eq.~(\ref{betaint}) is symmetric in $\beta$ it is enough to
use $\beta_1$ which is in the interval $[0,\pi]$ and multiply by 2.
Thus we end up with
\begin{equation}
\int_{0}^{2 \pi} d \beta\, \delta(f(\beta)) = \frac{2}{\left| 
\frac{df(\beta)}{d \beta} \right| _{\beta = \beta_{1}}} 
\Theta \left( \left| \frac{df(\beta)}{d \beta} \right|_{\beta = 
\beta_{1}}^{2}\right).
\end{equation}
The derivative may be written in the form
\begin{equation}
\left| \frac{df(\beta)}{d \beta} \right| _{\beta = \beta_{1}} = 
\sqrt{a z^2 + b z +c},
\end{equation}
where $z\equiv\cos\alpha$ and
\begin{eqnarray}
a & = & p_2^2 (-p_1^2 - p_3^2 + 2 p_1 p_3 \cos \theta), 
\nonumber\\
b & = & 2 \omega m_N p_1 p_2 - 2 p_1 p_2 p_3^2 - 2 \omega m_N p_2 p_3 
\cos \theta + 2 p_1^2 p_2 p_3 \cos \theta \nonumber \\
& & +\, 2 p_2 p_3^3 \cos \theta - 2 p_1 p_2 p_3^2 \cos^2 \theta, 
\nonumber\\
c & = & \omega^2 m_N^2 + 2 \omega m_N p_3^2 + p_2^2 p_3^2 - p_3^4 - 
2 \omega m_N p_1 p_3 \cos \theta \nonumber \\ 
& & +\, 2 p_1 p_3^3 \cos \theta - p_1^2 p_3^2 \cos^2\theta - 
p_2^2 p_3^2 \cos^2\theta.
\end{eqnarray}
The integration of the $\delta$ function also fixes
$E_4=E_1+E_2-E_3+\omega$ which is used to determine the Pauli-blocking
factor of particle 4 in the phase-space integral.

In order to perform the integration over $dz=d\cos\alpha$ analytically
we note that partial integrations allow us to reduce the integrand to
a sum of two terms. One of them is independent of $z$ while the other
is of the form $1/(qz + r)$ where $q$ and $r$ are expressions which do
not depend on $z$. We find explicitly
\begin{eqnarray}
\int_{-\infty}^{+\infty} \frac{dz}{\sqrt{az^{2}+bz+c}}\, 
\Theta (az^{2}+bz+c) & = & \frac{\pi}{\sqrt{-a}}\,
\Theta (b^{2}-4ac),\nonumber\\
\int_{-\infty}^{+\infty} \frac{dz}{\sqrt{az^{2}+bz+c}}\, 
\frac{1}{qz+r}\, \Theta(az^{2}+bz+c) 
& = & \frac{\pi\,\Theta (b^{2}-4ac)}
{\sqrt{-a r^2 q^{-2}+b r q^{-1}-c}}.
\label{eq:funda}
\end{eqnarray}
The step function in the integrand singles out the physically relevant
range where $-1\leq z\leq+1$.

After these analytic manipulations we are left with a four dimensional
numerical integral over $dp_1 dp_2 dp_3d \cos \theta$ where the
integration region is determined by the step function $\Theta
(b^{2}-4ac)$.



\newpage

\begin{figure}
\epsfxsize=10cm
\hbox to\hsize{\hfill\epsfbox{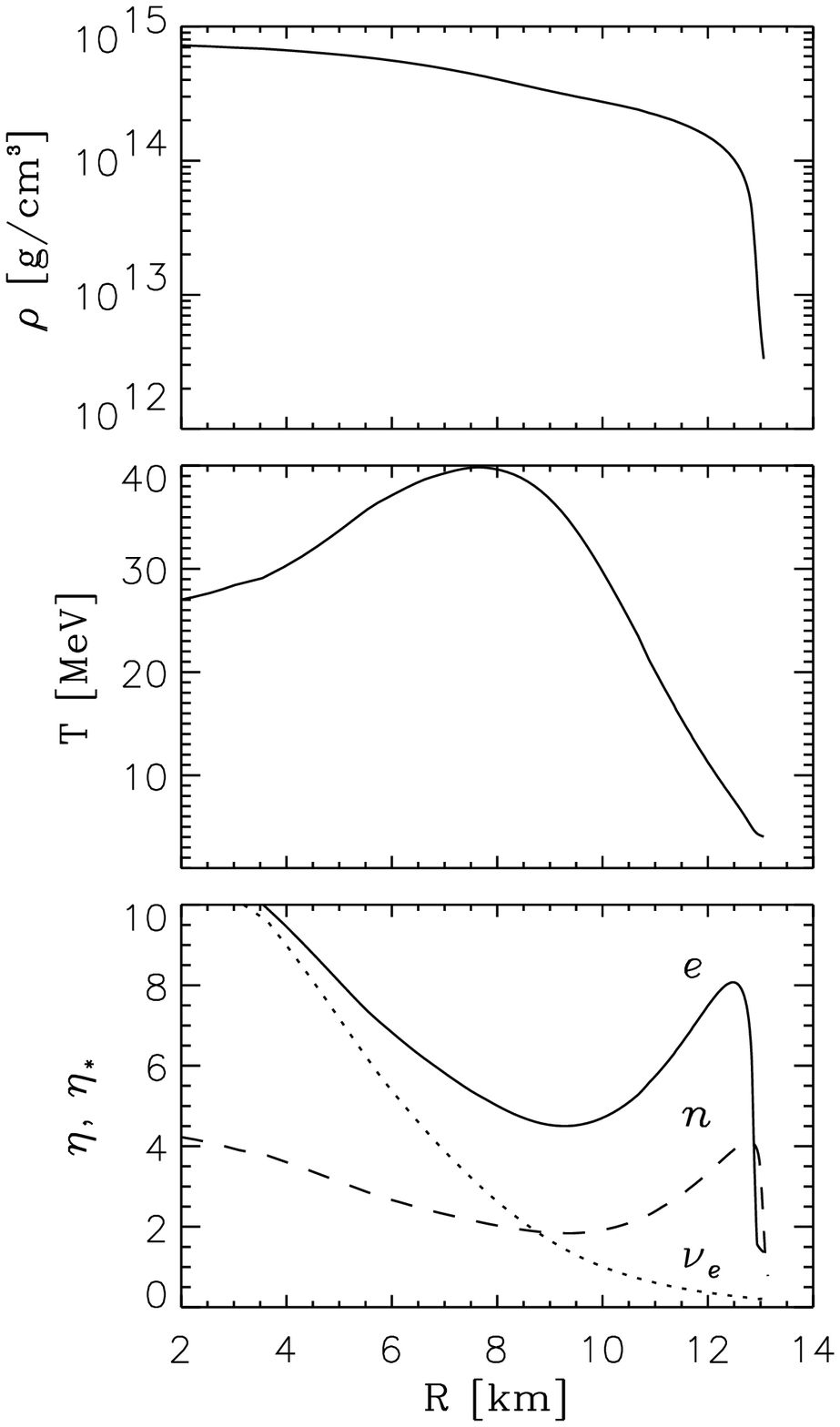}\hfill}
\vskip1cm
\figcaption{Physical parameters of the proto neutron star
model S2BH\_0 of Keil, Janka \& Raffelt (1995) which represents a 
SN core $1\,\rm s$ after bounce. For the neutrons we show the 
effective degeneracy parameter $\eta_*$ as defined in 
Eq.~(\ref{eq:effdeg}), taking the vacuum nucleon
mass for~$m_N$.} 
\end{figure}

\begin{figure}
\epsfxsize=11cm
\hbox to\hsize{\hfill\epsfbox{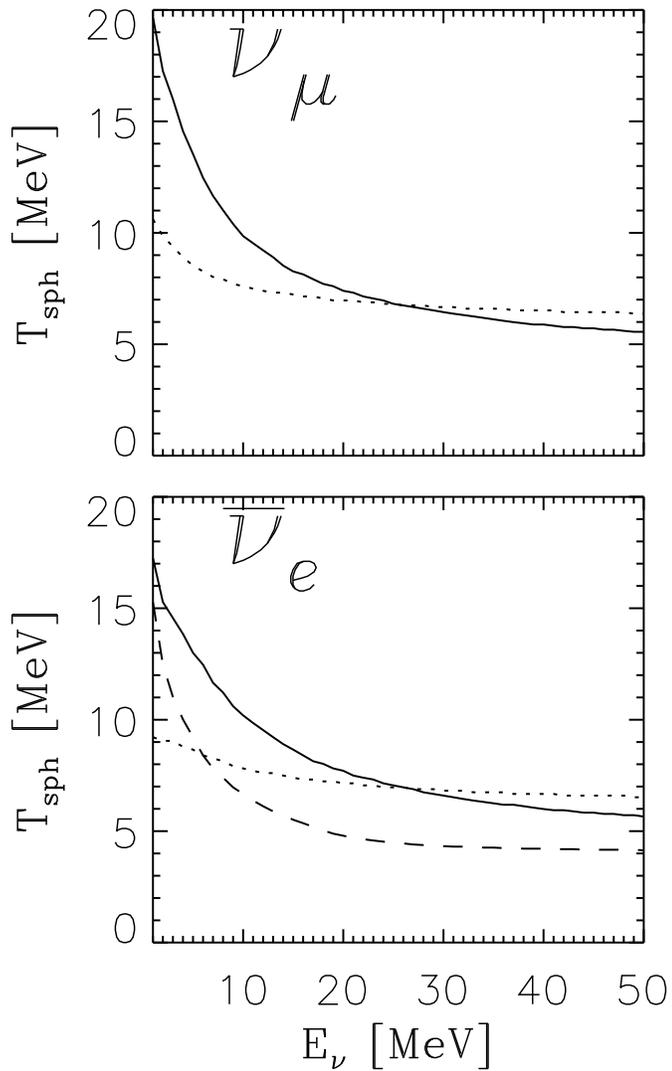}\hfill} 
\vskip1.5cm
\figcaption{Temperature of the medium (SN model of Fig.~1) at last
energy exchange for different processes:  Scattering on $e^{\pm}$
(solid line), inverse bremsstrahlung $\nu\bar\nu nn\to nn$ (short
dashes), and $\beta$-absorption $\bar{\nu}_e p \to n e$ (long
dashes).}
\end{figure}

\begin{figure}
\hbox to\hsize{\hfill\epsfbox{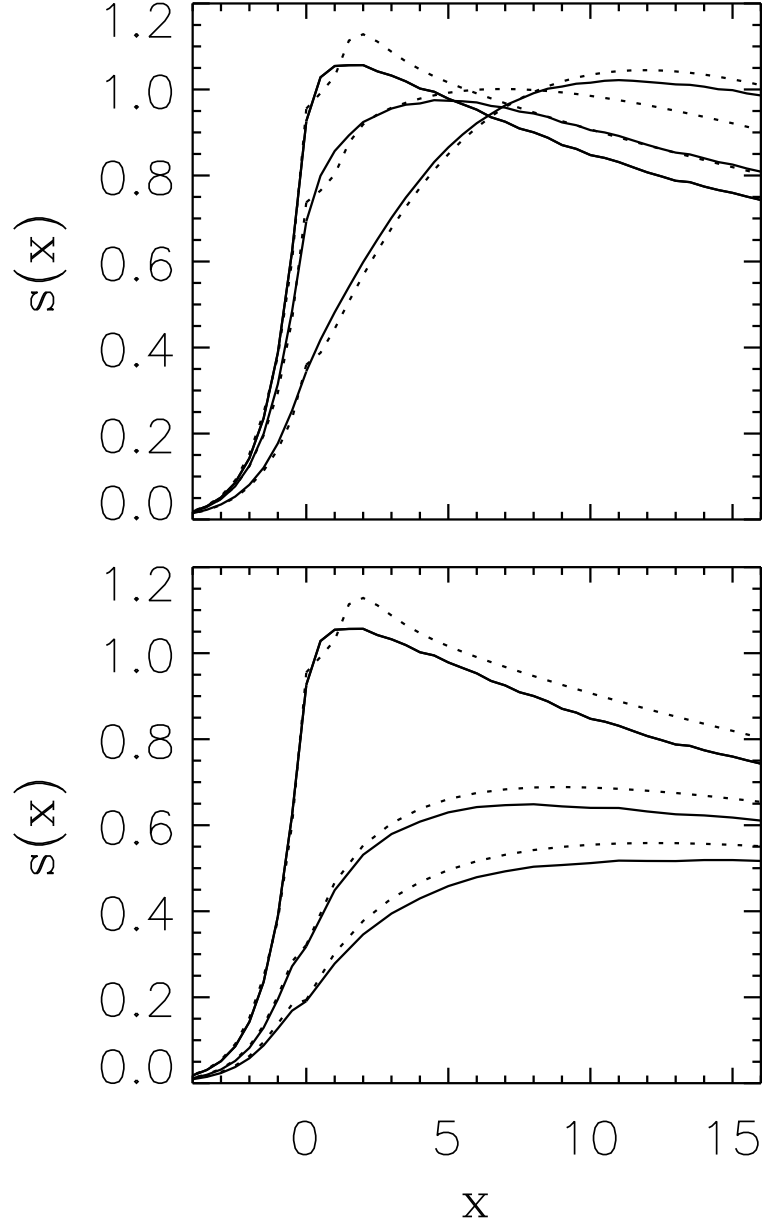}\hfill} 
\vskip1.5cm 
\figcaption {Dimensionless scattering kernel $s(x,y,\eta_*)$. 
{\it Upper Panel:} Pion mass ignored ($y=0$) with $\eta_{\ast}=0.31$,
1.01, and 2.42 from top to bottom, corresponding to $\eta=-2$, 0, and
2, respectively.  The solid curves are numerical results whereas the
short-dashed ones are the analytical approximation Eq.~(\ref{analyt}).
{\it Lower Panel:} 
Nondegenerate case ($\eta=-2$, $\eta_{\ast} = 0.31$)
with $y=0$, 2, and 4 from top to bottom.}
\end{figure}

\begin{figure}
\hbox to\hsize{\hfill\epsfbox{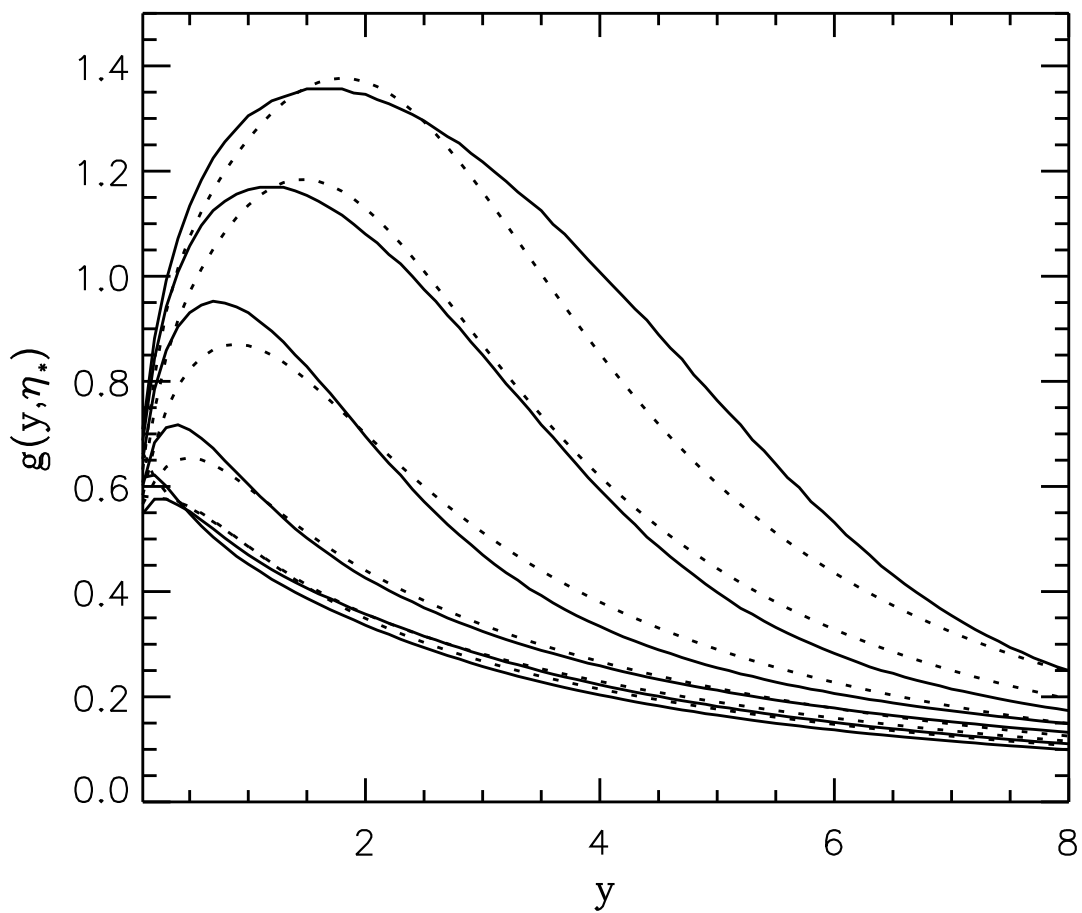}\hfill} \vskip1.5cm
\figcaption{Function $g(y,\eta_*)$ as defined in Secs.~4.1 and 4.5 to
take multiple-scattering effects into account. From bottom to top
the curves are for $\eta_*=0.31$, 1.01, 2.42, 4.22, 6.14, and 8.11
respectively.  The numerically
calculated $g(y,\eta_*)$ are shown as full lines, whereas the 
analytic fit from Eq.\ (\ref{multia}) is represented by the dotted
lines.}
\end{figure}

\begin{figure}
\hbox to\hsize{\hfill\epsfbox{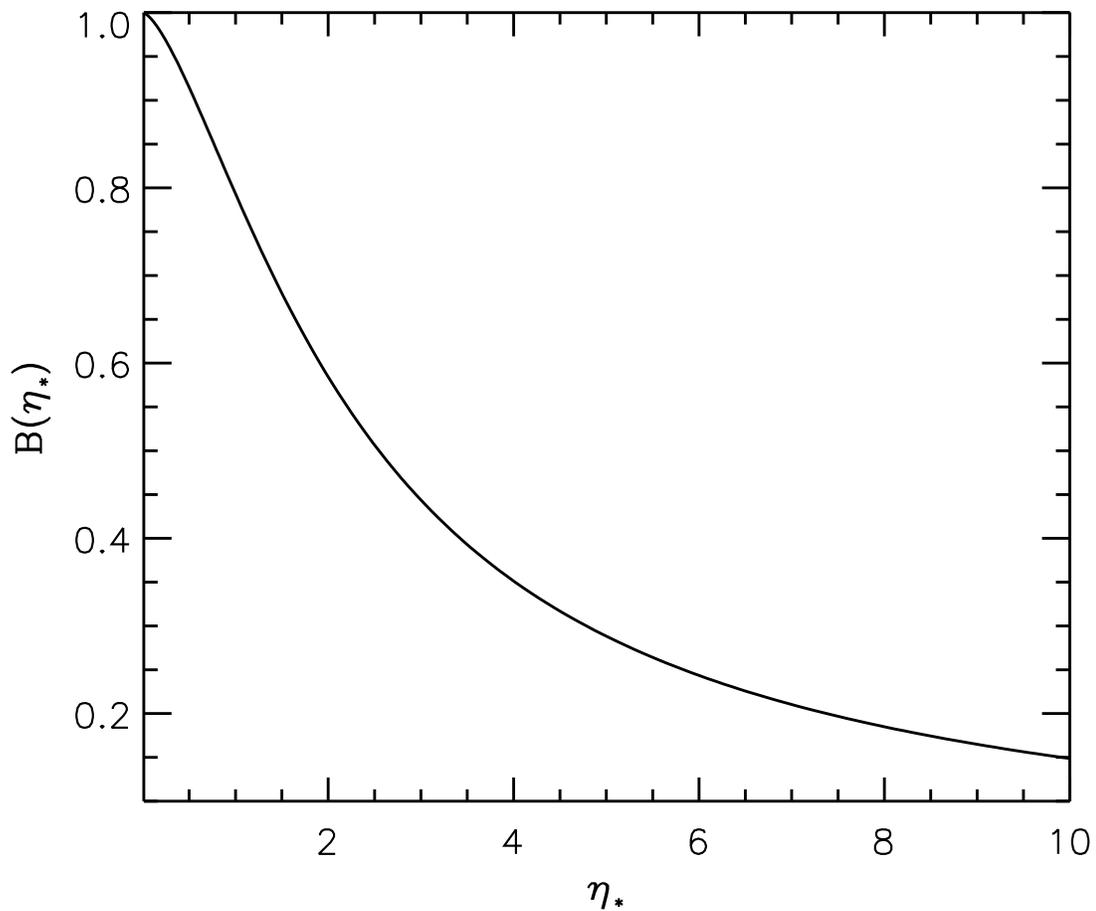}\hfill} \vskip1.5cm
\figcaption{The normalization function
$B(\eta_{\ast})$ defined in Eq.~(\ref{norm}) as
a function of the nucleon degeneracy parameter $\eta_{\ast}$.}
\end{figure}


\newpage

\begin{deluxetable}{lcccccccccc}
\footnotesize
\tablecaption{Neutrino Energy Spheres}
\tablewidth{0pt}
\tablehead{Flavor:& \multicolumn{5}{c}{$\nu_{\mu}$, $\nu_{\tau}$,
$\bar\nu_{\mu}$, $\bar\nu_{\tau}$} 
& \multicolumn{5}{c}{$\bar{\nu}_{e}$}\nl
$f_{\rm brems}$ &0&1&5&10&20&0&1&5&10&20}
\startdata
$\rho_{\rm sph}$ [$10^{14}$ g/cm$^3$] & 
1.20&0.99&0.85&0.78&0.71&0.82&0.77&0.70&0.66&0.62\nl
$T_{\rm sph}$ [MeV] & 
8.78&7.50&6.67&6.31&5.99&6.52&6.25&5.94&5.77&5.60\nl
$T^{\infty}_{\rm sph}$ [MeV]& 
7.39&6.32&5.63&5.33&5.06&5.49&5.25&5.02&4.87&4.73\nl
$\tau_{\rm sph}^{\rm tot}$ & 
16.3&7.11&4.42&3.49&2.83&3.83&2.78&2.18&1.93&1.76\nl
\enddata
\end{deluxetable}

\begin{deluxetable}{rccccc}
\footnotesize
\tablecaption{$s(x,y,\eta_*)$ in the form
``Numerical Value/Analytical Approximation''.}
\tablewidth{0pt}
\tablehead{\multicolumn{1}{l}{$y={}$}& 0 & 2 & 4 & 6 & 8}
\startdata
\multicolumn{6}{c}{$\eta=0$ \quad ($\eta_{\ast} = 1.01$)} \nl \hline
$x=0$&0.701/0.735&0.267/0.277&0.166/0.171&0.117/0.117&0.088/0.086\nl 
1&0.861/0.799&0.389/0.412&0.248/0.272&0.177/0.193&0.134/0.145\nl 
2&0.926/0.916&0.482/0.508&0.321/0.352&0.233/0.259&0.179/0.198\nl 
3&0.959/0.955&0.551/0.579&0.381/0.416&0.284/0.314&0.222/0.246\nl 
4&0.973/0.977&0.600/0.633&0.429/0.466&0.327/0.360&0.259/0.287\nl
5&0.976/0.991&0.636/0.673&0.467/0.506&0.362/0.398&0.291/0.322\nl
6&0.971/0.998&0.660/0.702&0.496/0.538&0.391/0.429&0.319/0.351\nl
7&0.959/1.000&0.675/0.724&0.518/0.562&0.414/0.454&0.341/0.375\nl
8&0.945/0.998&0.685/0.738&0.534/0.581&0.433/0.474&0.360/0.395\nl
9&0.928/0.993&0.690/0.748&0.546/0.595&0.448/0.489&0.376/0.411\nl
10&0.911/0.985&0.691/0.753&0.554/0.605&0.459/0.501&0.389/0.424\nl
\hline\hline
\multicolumn{6}{c}{$\eta=2$ \quad ($\eta_{\ast} = 2.42$)} \nl 
\hline
$x=0$&0.349/0.360&0.165/0.172&0.111/0.115&0.082/0.084&0.064/0.064\nl 
1&0.485/0.443&0.252/0.270&0.172/0.190&0.128/0.141&0.101/0.110\nl 
2&0.600/0.567&0.342/0.368&0.240/0.267&0.182/0.203&0.144/0.160\nl 
3&0.705/0.676&0.431/0.463&0.310/0.343&0.238/0.266&0.191/0.212\nl 
4&0.794/0.770&0.513/0.550&0.378/0.415&0.295/0.326&0.238/0.263\nl 
5&0.867/0.847&0.585/0.624&0.440/0.478&0.349/0.381&0.285/0.311\nl 
6&0.924/0.909&0.646/0.685&0.495/0.532&0.397/0.429&0.328/0.353\nl 
7&0.964/0.957&0.694/0.735&0.541/0.577&0.439/0.469&0.366/0.390\nl 
8&0.993/0.992&0.733/0.773&0.580/0.613&0.476/0.502&0.400/0.421\nl 
9&1.011/1.016&0.763/0.801&0.611/0.641&0.506/0.529&0.428/0.446\nl 
10&1.020/1.032&0.784/0.821&0.636/0.663&0.531/0.550&0.453/0.466\nl 
\hline\hline
\multicolumn{6}{c}{$\eta=4$ \quad ($\eta_{\ast} = 4.22$)} \nl 
\hline
$x=0$&0.153/0.154&0.086/0.087&0.063/0.064&0.049/0.050&0.040/0.040\nl 
1&0.229/0.219&0.136/0.145&0.100/0.109&0.078/0.086&0.064/0.070\nl 
2&0.313/0.312&0.197/0.218&0.147/0.167&0.117/0.133&0.096/0.109\nl 
3&0.408/0.415&0.269/0.304&0.204/0.235&0.163/0.189&0.135/0.156\nl 
4&0.509/0.522&0.349/0.396&0.269/0.310&0.217/0.252&0.181/0.209\nl 
5&0.613/0.625&0.434/0.489&0.339/0.387&0.276/0.316&0.231/0.263\nl 
6&0.712/0.720&0.518/0.577&0.409/0.460&0.337/0.378&0.284/0.317\nl 
7&0.803/0.802&0.598/0.654&0.478/0.525&0.397/0.434&0.336/0.366\nl 
8&0.884/0.871&0.671/0.720&0.543/0.582&0.454/0.484&0.387/0.410\nl 
9&0.953/0.927&0.736/0.773&0.602/0.629&0.506/0.525&0.435/0.447\nl 
10&1.011/0.970&0.793/0.814&0.654/0.667&0.554/0.559&0.478/0.477\nl 
\hline
\tablebreak
\multicolumn{6}{c}{$\eta=6$ \quad ($\eta_{\ast} = 6.14$)} \nl 
\hline
$x=0$&0.074/0.074&0.046/0.046&0.036/0.036&0.029/0.029&0.025/0.025\nl 
1&0.114/0.116&0.074/0.080&0.057/0.063&0.047/0.052&0.040/0.044\nl 
2&0.164/0.177&0.111/0.127&0.087/0.101&0.072/0.084&0.061/0.071\nl 
3&0.227/0.253&0.159/0.189&0.126/0.152&0.104/0.126&0.088/0.107\nl 
4&0.300/0.340&0.217/0.265&0.173/0.214&0.144/0.179&0.123/0.152\nl 
5&0.385/0.435&0.284/0.350&0.229/0.284&0.192/0.238&0.165/0.203\nl 
6&0.476/0.530&0.359/0.438&0.293/0.359&0.247/0.301&0.212/0.258\nl 
7&0.572/0.621&0.440/0.526&0.361/0.432&0.306/0.364&0.264/0.313\nl 
8&0.669/0.705&0.522/0.606&0.432/0.501&0.368/0.424&0.319/0.364\nl 
9&0.763/0.779&0.603/0.678&0.503/0.563&0.431/0.477&0.375/0.411\nl 
10&0.851/0.841&0.681/0.738&0.572/0.615&0.492/0.523&0.431/0.452\nl 
\hline\hline
\multicolumn{6}{c}{$\eta=8$ \quad ($\eta_{\ast} = 8.11$)} \nl 
\hline
$x=0$&0.043/0.041&0.029/0.027&0.023/0.022&0.020/0.018&0.017/0.016\nl 
1&0.068/0.068&0.046/0.048&0.037/0.039&0.031/0.033&0.027/0.029\nl 
2&0.100/0.108&0.071/0.079&0.057/0.065&0.048/0.055&0.042/0.048\nl 
3&0.142/0.161&0.103/0.122&0.084/0.101&0.072/0.086&0.062/0.074\nl 
4&0.194/0.227&0.145/0.178&0.119/0.147&0.102/0.126&0.088/0.109\nl 
5&0.257/0.302&0.196/0.246&0.163/0.204&0.139/0.175&0.121/0.152\nl 
6&0.331/0.385&0.257/0.322&0.214/0.269&0.184/0.230&0.161/0.201\nl 
7&0.413/0.469&0.325/0.403&0.273/0.338&0.236/0.290&0.207/0.253\nl 
8&0.503/0.552&0.401/0.484&0.339/0.408&0.294/0.351&0.259/0.306\nl 
9&0.597/0.630&0.482/0.561&0.409/0.474&0.356/0.409&0.315/0.357\nl 
10&0.693/0.700&0.566/0.631&0.483/0.535&0.422/0.462&0.374/0.404\nl 
\enddata
\end{deluxetable}

\begin{deluxetable}{rccccc}
\footnotesize
\tablecaption{Dimensionless mean free path against pair-absorption
$\xi$ as defined in Eq.~(\ref{xidef}).
Values are given as ``without/with'' multiple scattering.}
\tablewidth{0pt}
\tablehead{\multicolumn{1}{l}{$\eta_\ast={}$}& 1 & 2 & 4 & 6 & 8}
\startdata
$y=1$&0.750/0.712&0.666/0.488&0.471/0.102&0.321/0.017&0.222/0.005\nl
2&0.609/0.604&0.557/0.524&0.408/0.237&0.284/0.052&0.199/0.013\nl
4&0.447/0.447&0.417/0.415&0.320/0.297&0.230/0.157&0.166/0.063\nl
6&0.345/0.345&0.327/0.327&0.259/0.256&0.192/0.179&0.141/0.112\nl
8&0.276/0.276&0.264/0.264&0.215/0.215&0.164/0.161&0.123/0.116\nl

\enddata
\end{deluxetable}

\end{document}